\documentclass[useAMS,usenatbib,onecolumn]{mn2e}
\usepackage{natbib}
\usepackage{graphicx}
\usepackage{times}
\usepackage{rotate}
\usepackage{color,url}
\newcommand{\bmf}[1]{\mathbf {#1}}

\newcommand{\simgt}{\lower.5ex\hbox{$\; \buildrel > \over \sim \;$}}
\newcommand{\simlt}{\lower.5ex\hbox{$\; \buildrel < \over \sim \;$}}
\newcommand{\ave}[1]{\left\langle #1\right\rangle}
\newcommand{\rmd}{\ensuremath{\mathrm{d}}}

\begin{document}
\title[Where are the LRGs in their host halos?]
{
Where are the Luminous Red Galaxies (LRGs)? 
Using correlation measurements and lensing to relate LRGs to dark matter halos
}

\author[Hikage, Mandelbaum, 
Takada \& Spergel]{Chiaki Hikage$^1$, 
Rachel Mandelbaum$^{2,3}$,
Masahiro Takada$^4$, 
and
David N. Spergel$^{2,4}$ \\
$^1$ Kobayashi Maskawa Institute (KMI), Nagoya University, Aichi 
464-8602, Japan\\
$^2$ Department of Astrophysical Sciences, Princeton University, Peyton
Hall, Princeton NJ 08544, USA \\
$^3$ Department of Physics, Carnegie Mellon University, Pittsburgh, PA 15213, USA\\
$^4$ Kavli Institute for the Physics and Mathematics of the Universe
(Kavli IPMU, WPI), 
The University of Tokyo, Chiba 277-8582, Japan}
\maketitle

\label{firstpage}

\begin{abstract}
Nonlinear redshift-space distortions, the Finger-of-God (FoG) effect, 
can complicate the
  interpretation of the galaxy power spectrum.  Here, we demonstrate
  the method proposed by \cite{Hikageetal:12} to use complimentary
  observations to directly constrain this effect on the data.
We use catalogs of Luminous Red Galaxies (LRGs) and
photometric galaxies from the Sloan Digital Sky Survey (SDSS) Data Release 7 (DR7) to measure the redshift-space power
spectrum of LRGs, the cross-correlation of LRGs with the shapes of
background photometric galaxies (galaxy-galaxy weak lensing), and the
projected cross-correlation of LRGs with photometric galaxies having
similar photometric redshifts to the LRG spectroscopic redshift. All of  these measurements use
 a reconstructed halo field.
While we use the position of each LRG for single LRG
systems, we compare the measurements using different halo-center proxies
for multiple-LRG systems (4.5 per cent of all the halos): 
the brightest LRG position (BLRG), the faintest LRG position (FLRG) and
their 
arithmetical
mean 
position (Mean), respectively, in each system. We
find significant differences in the measured correlations of different
centers, showing consistent off-centering effects in the three
observables. By comparing the measurements with a halo model that
treats the satellite photometric galaxies as being distributed according to a
generalized NFW profile, we find
that $\sim40$ (70) per cent of BLRGs (FLRGs) are off-centered satellite 
galaxies in the multiple-LRG systems. The satellite LRGs have
typical off-centering radius of $\sim$400~kpc/$h$, 
and velocity dispersion of about $500$~km$/$s
in host halos with a 
mean mass of $1.6\times 10^{14}~M_\odot/h$. 
We show that, if LRGs in the single LRG systems have 
similar offsets, the residual FoG contamination in the LRG power
spectrum can be significant at $k\simgt 0.1~h/{\rm Mpc}$, which may
cause a bias in cosmological parameters determined by the shape of the power spectrum, such as the neutrino mass.  Our results demonstrate that overlapping spectroscopic and imaging galaxy
surveys can be used to observationally calibrate the FoG contamination and more robustly use the galaxy power spectrum for cosmological measurements.
\end{abstract}


\section{Introduction}
\label{sec:intro}
Large-scale structure surveys are one of the primary tools for studying the nature of dark matter, 
dark energy and the initial conditions in the early universe    
\citep{DavisHuchra:82,deLapparentetal:86,Kirshneretal:87,SDSS,Peacocketal:01}.
Over the coming decade, astronomers are embarking on  even larger
surveys including 
the Baryon Oscillation Spectroscopic Survey (BOSS\footnote{\url{http://cosmology.lbl.gov/BOSS/}}, \citealt{2012arXiv1208.0022D}),
WiggleZ\footnote{\url{http://wigglez.swin.edu.au/site/}}
\citep{Blakeetal:11}, Vipers\footnote{\url{http://vipers.inaf.it/}},
FMOS\footnote{\url{http://www.naoj.org/Observing/Instruments/FMOS/}},
HETDEX\footnote{\url{http://hetdex.org/}},
BigBOSS\footnote{\url{http://bigboss.lbl.gov/}} \citep{BigBOSS},
Subaru
Prime Focus Spectrograph (PFS\footnote{\url{http://sumire.ipmu.jp/pfs/intro.html}}; \citealt{Ellisetal:12}),
Euclid\footnote{\url{http://sci.esa.int/euclid}}, and
WFIRST\footnote{\url{http://wfirst.gsfc.nasa.gov/}}.  The upcoming
generation of surveys is motivated by our desire to understand cosmic
acceleration and to measure the composition of the universe by
simultaneously measuring geometry and dynamics
\citep[][]{WangSpergelStrauss:99,Eisensteinetal:99,Tegmarketal:04,Coleetal:05}. 
Measurements of the baryon acoustic oscillation (BAO) scale provide us
with a robust geometrical probe of the angular diameter distance and
the Hubble expansion rate
\citep{Eisensteinetal:05,Percivaletal:07BAO,Andersonetal:12}.  Observations of
redshift-space distortion measure the growth rate of structure
formation \citep[][]{Zhangetal:07,Guzzoetal:08,Wang:08,Guziketal:10,
  Whiteetal:09,PercivalWhite:09,SongPercival:09,SongKayo:10,Yamamotoetal:10,Tangetal:11,Reidetal:12}.
Combining measurements of 
the
geometry of the universe and
the growth of structure formation 
provides a key clue to understanding the
nature of dark energy as well as constraining
 properties of gravity on cosmological scales
\citep[][]{DETF,Peacocketal:06}.

If the distribution of galaxies in redshift space simply traced the underlying distribution of
matter in real space, then cosmologists could easily interpret the large-scale structure
observations.  The universe, however, is not so simple:  galaxies do not necessarily sit
in the centers of dark matter halos and correspondingly, they are moving in these halos.
This  Finger-of-God
(FoG) effect \citep[][]{Jackson:72,PeacockDodds:94,Scoccimarro:04}
is a significant source of systematic uncertainty in the cosmological
  interpretation of the redshift-space galaxy power spectrum

\cite{Reidetal:09} advocated using halos rather than Luminous Red
Galaxies \citep[LRGs;][]{Eisensteinetal:01} to  mitigate FoG effects.  In an analysis of LRGs from the Sloan Digital Sky
Survey\footnote{\url{http://www.sdss.org/}} (SDSS) Data Release 7 (DR7) catalog,
\cite{Reidetal:10} implemented this scheme by removing satellite LRGs
from within halos hosted by 
LRGs, with the aid of mock catalogs and the halo model
prescription. 
Once
such a halo catalog is constructed, the clustering properties of halos are
easier to model, because halos have only bulk motions in large-scale
structure, and therefore have a reduced FoG effect.
However, even if one can perfectly choose one galaxy in each halo 
(e.g., the brightest LRG or the brightest cluster galaxy), 
there still remains uncertainty;
the position of the chosen galaxy is not necessarily at the halo center
\citep{Skibbaetal:11}, and the off-centered galaxies have an internal
motion relative to the true halo center, which  causes residual
FoG contamination 
\citep{Seljak:01,White:01,Hikageetal:12}.

In our previous paper \citep{Hikageetal:12}, we developed a method to
correct the residual FoG effect by combining the redshift-space galaxy
power
spectrum with the cross-correlation of spectroscopic galaxies with
images of background galaxies -- the so-called galaxy-galaxy weak lensing
\citep[e.g., ][]{Mandelbaumetal:06,Sheldonetal:09,Leauthaudetal:10,Okabeetal:10}.
The galaxy-galaxy lensing measures the average mass distribution around
spectroscopic galaxies; i.e., at small scales, it reveals the mass profile of the halos hosting
the lens galaxies, while on large scales it is sensitive to the mass distribution surrounding the host halos. However, if we include off-centered galaxies and
use their positions as a proxy for the center of each halo, the lensing signals at angular scales smaller than the
typical offset scale are diluted. In other words, comparing the
galaxy-galaxy lensing signals of different halo-center proxies can be
used to infer 
the amount of the off-centered
galaxy contamination
\citep[][also see \citealt{OguriTakada:11} 
for a useful formulation of the off-centering effect
on galaxy-galaxy weak lensing]{Johnstonetal:07,Leauthaudetal:10,Okabeetal:10,Georgeetal:12}. 
Hence, a weak-lensing based calibration of the FoG effect in
redshift-space power spectrum measurements can be feasible if
spectroscopic and imaging surveys observe the same region of the
sky. Fortunately, this is the case for 
many upcoming surveys: the BOSS and the Subaru Hyper Suprime-Cam (HSC) Survey\footnote{\url{http://www.naoj.org/Projects/HSC/index.html}}
\citep[][]{Miyazakietal:06}, the Subaru PFS and HSC surveys (Subaru
Measurements of Images and Redshifts: the SuMIRe project), Euclid and
WFIRST or a combination of LSST \citep{LSST} with spectroscopic surveys.

The goal of this paper is to implement this method and to determine how
LRGs are distributed in dark matter halos.  For this paper,
we use
the SDSS DR7 catalog to constrain the fraction of off-centered LRGs and
the amount of the residual FoG contamination in the LRG power
spectrum measurement.  
We focus on 
systems that contain multiple LRGs in the 
same halo, defined based on the friend-of-friends algorithm
\citep{ReidSpergel:09,Reidetal:10}. Then, we compare the redshift-space
power spectra and the cross-correlations of LRGs with shapes of
background photometric galaxies, measured using different halo-center
proxies. Furthermore, we also use the projected cross-correlations of
LRGs with photometric galaxies that have similar photometric redshifts
to the LRG redshift, and then compare the measurements using  
different halo centers.  We employ the following three proxies for the
halo center: the brightest LRG
position, 
 the faintest LRG
position, and the 
arithmetical
 mean position of LRGs (their 
center-of-mass position without any weighting)
in each
multiple-LRG system. 
We show that comparing the measurements to the halo model allows us to
constrain the population of off-centered LRGs, the amount of
off-centering effects, and the residual FoG contamination.

The structure of this paper is as follows. In Section~\ref{sec:model}, we
develop a halo model to compute the redshift-space power spectrum,
LRG-galaxy lensing and LRG-photometric galaxy cross-correlation
including off-centering effects. In Section~\ref{sec:results}, we show
the power spectrum, galaxy-galaxy lensing and the photometric galaxy cross-correlation
measured from the SDSS DR7 catalog of LRGs and photometric galaxies,
using the different halo-center proxies. Then we constrain the model
parameters by comparing the halo model predictions to the
measurements. Based on these findings, in Section~\ref{sec:fog_residual},
we discuss the residual FoG contamination in the LRG power
spectrum, and its impact on cosmological
parameters. Section~\ref{sec:conclusion} is devoted to discussion and
conclusions. Unless explicitly stated otherwise, throughout this paper
we assume a WMAP-normalized $\Lambda$CDM model as our fiducial cosmological
model \citep{Komatsu:09}: $\Omega_{\rm b}h^2=0.0226$, $\Omega_{\rm
cdm}h^2=0.1109$, $\Omega_\Lambda=0.734$ $\tau=0.088$,
$n_s=0.963$, and $\sigma_8 = 0.817$, where
$\Omega_{\rm b}$, $\Omega_{\rm cdm}$ and $\Omega_{\Lambda}$ are the
energy density parameters of baryon, CDM and dark energy (the
cosmological constant with $w_0=-1$ here), $\tau$ is the optical depth
to the last scattering surface, and $n_s$ 
is the tilt 
of the primordial curvature power spectrum.

\section{The effects of off-centered LRGs on the galaxy clustering and
 weak lensing correlations}
\label{sec:model}

To constrain the radial distribution of LRGs within their
host halos, we use three observables that
can be measured from the SDSS DR7 data \citep{SDSSDR7}: the
angle-averaged, redshift-space power
spectrum of LRGs, the weak lensing correlation of LRGs with background
galaxy images, and the projected cross-correlation of LRGs with photometric
redshift galaxies that have photo-$z$'s similar to the LRG redshift.
When 
computing these correlation functions, we use the same LRG catalog, but
compare the correlation functions measured using different 
halo-center proxies (see below).
 Before describing the measurements, we first 
describe our halo model for the correlation functions, 
including the effect of off-centered LRGs \citep[also see][]{OguriTakada:11,Hikageetal:12}.

\subsection{Connecting LRGs to halos and 
the radial offset distribution of LRGs}

Weak lensing studies \citep[e.g., ][]{Mandelbaumetal:06,Johnstonetal:07}
and clustering analyses
\citep[][]{Rossetal:07,Rossetal:08,Wakeetal:08,Zhengetal:09,ReidSpergel:09,Whiteetal:11}
have shown that most LRGs reside in massive halos of a few
$\times 10^{13}M_\odot$.  
While the typical
massive halo contains only one LRG, about 5 per cent of LRGs are found to
be satellite galaxies in a halo; there is a population of halos that
contain multiple LRGs \citep{ReidSpergel:09}.
Including these satellite galaxies in the correlation function
measurements alters the clustering signals at small scales. 
 This 
contribution complicates the estimation of cosmological parameters,
because the small-scale signals involve complicated physical processes
of galaxy formation/evolution that are still very challenging to
accurately model. Hence, in this paper, we employ a method similar to
that 
developed in \cite{ReidSpergel:09}  to
connect the distribution of LRGs to that of host halos by removing
satellite LRGs from the catalog \citep[see also][]{Reidetal:10}. In other words, we assume that we can
identify one LRG (or one halo-center proxy)
in each LRG system. 

Since we are interested in the LRG-related correlation
functions, which are statistical quantities,
we need to model 
the {\em average} radial distribution of LRGs within host halos 
 as a function of halo mass $M$:
$p_{\rm off}(r)$, which is normalized as
$4\pi\int^{r_{\rm vir}}_0 r^2\rmd r \,p_{\rm off}(r)=1$.  
If LRGs are at the halo center, $p_{\rm
off}(r)=\delta_D(r)/4\pi r^2$.
For off-centered LRGs, 
the Fourier
transform of the radial profile is given as
\begin{equation}
\tilde{p}_{\rm off}(k; M)=4\pi\int^{r_{\rm vir}}_0 r^2\rmd r \, p_{\rm
 off}(r)\frac{\sin kr}{kr},
\label{eq:poff} 
\end{equation}
where we will often omit the halo mass dependence of $p_{\rm off}$ for
notational simplicity.   For central LRGs, $\tilde{p}_{\rm off}(k)=1$. 
In this paper, we assume for simplicity that offset LRGs follow a Gaussian distribution,
$p_{\rm off}(r)\propto \exp[-r^2/2R_{\rm off}^2]$, and 
treat $R_{\rm off}$ as a free parameter to constrain from the
measured correlation functions.  However, the exact shape of the $p_{\rm off}( r)$ is less important than
its characteristic width.

\subsection{The FoG effect on the redshift-space power spectrum of LRGs}

As we discussed in detail in \cite{Hikageetal:12}, the major effect of
off-centered LRGs on the angle-averaged, redshift-space power spectrum
of LRGs is the nonlinear
redshift-space distortion effect, the so-called Finger-of-God (FoG) effect. 
If an LRG is offset from the true halo center, the LRG should be in 
motion with respect to it.  This internal motion within halos causes
the FoG effect on the LRG power spectrum, and can occur even if
selecting one LRG (or one center more generally) 
per halo, provided that the selected LRG or center is not the
center of mass of the halo. By employing the halo model approach, we
can model the redshift-space LRG power spectrum at a redshift
$z$ as
\begin{eqnarray}
P_{s,{\rm LRG}}(k,\mu)&=&\left[\frac{1}{\bar{n}_{\rm
LRG}}\int\!\rmd M~\frac{\rmd n}{\rmd M}b(M)N_{\rm HOD}(M)\tilde{p}_{
s, {\rm off}}(k,\mu; M) \right]^2P^{\rm NL}_{s,{\rm m}}(k,\mu),
\label{eq:ps_lrg}
\end{eqnarray}
where $\mu$ is the cosine of the angle between the line-of-sight direction and
the wavevector $\bmf{k}$, i.e.  $\mu\equiv k_{\parallel}/k$;
$\rmd n/\rmd M$ is
the halo mass function of halos with mass $M$ and at redshift $z$;
$b(M)$ is the linear halo bias function;
$N_{\rm HOD}(M)$ is the halo
occupation distribution (HOD) specifying the probability that halos of
mass $M$ host LRGs  ($N_{\rm HOD}\le 1$ in
our setting); $\bar{n}_{\rm LRG}$ is the mean number density of LRGs
defined as $\bar{n}_{\rm LRG}\equiv \int\!\rmd M (\rmd
n/\rmd M) N_{\rm HOD}(M)$;
$P_{s,{\rm m}}^{\rm NL}$ is the nonlinear redshift-space power spectrum
of dark matter. The offset function $\tilde p_{s,{\rm off}}(k,\mu)$ is
defined by convolving the real-space offset profile $p_{\rm off}$
(Eq.~\ref{eq:poff}) with the velocity function along the line-of-sight
direction via the virial theorem. See \cite{Hikageetal:12} for details.
Although we employ the linear halo bias in the above equation, which is not
a good approximation for a high-precision measurement in the quasi-nonlinear scales \citep{Baldaufetal:10}, this is not crucial for the
following results, because we will focus on fractional differences between the 
redshift-space power spectra of
different halo center proxies, where the bias factor or the dark matter
power spectrum cancel out in the fractional difference. 

To make the model more general, we introduce an additional
parameter $q_{\rm cen}$ ($0\le q_{\rm cen}\le 1$) to represent the 
fraction of central LRGs among all the LRGs in the sample: the fraction
of central LRGs is 
$q_{\rm cen}$, while $1-q_{\rm cen}$ is the fraction of satellite
LRGs. 
In this case, we can
model the resulting redshift-space power spectrum by replacing
$\tilde{p}_{\rm s,{\rm off}}$ in Eq.~(\ref{eq:ps_lrg}) as
\begin{equation}
\tilde{p}_{s,{\rm off}}(k,\mu)\longrightarrow
q_{\rm cen} + (1-q_{\rm cen}) \tilde{p}_{s,{\rm off}}(k,\mu). 
\label{eq:qcen_fog}
\end{equation}
Here we assumed that, for the first term, the central galaxies have no FoG
suppression, i.e. are at rest at the center of the host halo, so that
their 
$\tilde{p}_{s, {\rm off}} = 1$.

In this paper, we  consider the angle($\mu$)-averaged power
spectrum defined as
\begin{equation}
\bar{P}_{s,{\rm LRG}}(k)=\int^{1}_{-1}\frac{d\mu}{2}P_{s,{\rm
 LRG}}(k,\mu). 
\label{eq:ps_ave}
\end{equation}
For notational simplicity, we will hereafter omit the notation
``$\bar{\hspace{1em} }$'' of $\bar{P}_{s,{\rm
LRG}}(k)$.

\subsection{The effect of off-centered LRGs on LRG-galaxy weak
  lensing }
\label{sec:wl}

By cross-correlating the positions of LRGs with shapes of background
galaxies, we can measure the average {\em dark matter} distribution
around LRGs -- the so-called LRG-shear cross-correlation or LRG-galaxy
weak lensing
\citep{Mandelbaumetal:06,Mandelbaumetal:12}. 
If we include
off-centered LRGs in the analysis, they dilute the measured dark matter
profile at scales smaller than the off-centering radius
\citep{Johnstonetal:07,OguriTakada:11,Hikageetal:12}.  Under the assumption that the dark matter
halos are well-described by an NFW profile \citep{1996ApJ...462..563N}
or 
some other parametrized profile, 
we can use the measured surface density profile to
infer the distribution of off-centered halos.

For the observable of the LRG-galaxy weak lensing, 
we use the projected mass profile measured as a function of the projected
radius:
\begin{equation}
\Delta\Sigma(R)\equiv \int\!\frac{k\rmd k}{2\pi} C_{\Sigma g}(k) J_2(kR),
\label{eq:dsigma_def}
\end{equation}
where $J_2(x)$ is the 2nd-order Bessel function; $R$ is the projected
separation between LRG and background galaxy in units of the comoving
scale;
$C_{\Sigma g}(k)$ is the angular power spectrum of the projected mass
and LRG
cross-correlation, and  we give the power spectrum as a function of
the wavenumber $k$ for convenience in the rest of our discussion.
Here the projected radius needs to be estimated from the observed
angular separation for each LRG-galaxy pair on the sky: $R=\chi\theta$,
where $\chi$ is the comoving radial distance to each LRG. This
conversion requires a redshift for each lens LRG, which is
available from the SDSS DR7 catalog, and the assumption of a background
cosmological model, for which we assume the WMAP cosmology
\citep{Komatsuetal:10}. 
The $R$-average preserves
the physical scales of host halos, such as the virial radius. On the
other hand, if the cross-correlation is done in terms of the
  angular separation, then different scales are mixed for the
  halos at different redshifts, causing a smearing of the halo mass profile.

Employing the flat-sky approximation and the halo model,
we can express the power spectrum $C_{\Sigma
g}(k)$ as a sum of the 1- and 2-halo terms:
\begin{equation}
C_{\Sigma g}(k)=C^{1h}_{\Sigma g}(k)+C^{2h}_{\Sigma g}(k).
\end{equation}
 The 1-halo term  arises from the mass distribution within
one halo that hosts LRGs and is the dominant contribution to the signal
at small radii, while the 2-halo term arises from 
the mass distribution
surrounding the host halos and is the dominant contribution at large
radii.  Using Limber's approximation \citep{Limber:54}, the 1- and
2-halo terms are given as
\begin{eqnarray}
C_{\Sigma g}^{1h}(k)&\equiv &
\int\!\rmd\chi
f_{\rm LRG}(\chi)
\int\!\rmd M~\frac{\rmd n}{\rmd M}N^{\rm LRG}_{\rm HOD}(M; z) 
\left[
M\tilde{u}_{{\rm NFW}}\!\left(k; M,z
\right) \tilde{p}_{\rm off}(k; M)
+ m_{\rm sh, LRG}
\right], \nonumber\\
C^{2h}_{\Sigma g}(k)&\equiv& 
\int\!\rmd\chi~ f_{\rm LRG}(\chi)
\left[
\int\!\rmd M\frac{\rmd n}{\rmd M}b(M; z)N^{\rm LRG}_{\rm HOD}(M; z)
\right] \bar{\rho}_{\rm m0}P_m^L\!\left(k; z\right),
\label{eq:lens_2h}
\end{eqnarray}
%
where $\tilde{u}_{{\rm NFW}}(k; M,z)$ is the Fourier transform of an NFW
profile with mass $M$ at redshift $z$; $m_{\rm sh, LRG}$ is a
parameter representing the 
typical  mass of a sub-halo hosting a satellite LRG; $\bar{\rho}_{\rm m0}$ is the mean mass
density today; 
and $P^{\rm L}_m(k)$ is the linear
mass power spectrum \citep[see][for details]{Hikageetal:12}.  
Our use
of the 
mean mass density today ($\bar{\rho}_{\rm m0}$) rather than at the 
LRG redshift is due to our use of comoving units throughout this paper. 
The definition of the power spectra differs from the usual
definition by a factor of $\chi^{-2}$ \citep[e.g. see Eqs.~24 and 25 in][for
comparison]{Hikageetal:12}, arising from the conversion
$R\leftrightarrow \theta$ discussed above.
In the following analysis, we
will focus on the galaxy-galaxy lensing correlations at scales up to 
a few Mpc, where the 1-halo term is dominant. However, as a conservative
approach, we also include the 2-halo term treating the bias parameter as
a free parameter, and then derive constraints on the
off-centering 
parameters, marginalized over other parameter uncertainties. 
The function $f_{\rm
LRG}(\chi)$ in Eq.~(\ref{eq:lens_2h}) is the radial selection function of LRGs defined as
\begin{eqnarray}
f_{\rm LRG}(\chi)&\equiv& \frac{1}{\bar{n}^{\rm 2D}_{\rm LRG}}
 S_{\rm LRG}(z),\nonumber\\
\bar{n}_{\rm LRG}^{\rm 2D}&\equiv& 
\int\!\rmd\chi \,S_{\rm LRG}(\chi)
\int\!\rmd M\frac{\rmd n}{\rmd M}N^{\rm LRG}_{\rm HOD}(M), 
\label{eq:w_lrg}
\end{eqnarray}
where $S_{\rm LRG}(z)$ is the redshift selection function of LRGs,
$\bar{n}_{\rm LRG}^{\rm 2D}$ is the mean surface number density of LRGs, and
the selection function satisfies the normalization condition
$\int\!\rmd\chi f_{\rm LRG}(\chi) \int\!\rmd M(\rmd n/\rmd M)N_{\rm LRG}(M)=1$. 
For simplicity, we assumed that the sub-halo lensing contribution can be
characterized by a single number, a point mass, with a 
profile $\Delta \Sigma(R)= m_{\rm sh, LRG}/(\pi
R^2)$. See Section 3.2 in \cite{TakadaJain:03a} for the conversion relation between the
three-dimensional and projected mass density profiles in computing the
projected power spectrum in a halo model formulation. 

To gain insight into Eq.~(\ref{eq:lens_2h}), let us consider a case that 
the LRGs are in a thin redshift slice centered at $z_{\rm LRG}$, and
reside in halos with a narrow mass range centered at $\bar{M}$.
In this case, the power spectra can be
simplified as
\begin{eqnarray}
C_{\Sigma g}^{1h}(k)&\simeq &
\left[
\bar{M}\tilde{u}_{{\rm NFW}}\!\left(k; \bar{M},z_{\rm LRG}
\right) \tilde{p}_{\rm off}(k; \bar{M})
+ m_{\rm sh, LRG}
\right], \nonumber\\
C^{2h}_{\Sigma g}(k)&\simeq & 
b(\bar{M})\bar{\rho}_{\rm m0}P_m^L\!\left(k; z\right).
\label{eq:lens_2h_app}
\end{eqnarray}
These equations explicitly show that the 1-halo term arises form
the NFW halo profile, and the 2-halo term probes the power spectrum with
bias factor. 
When inverse Fourier-transforming the equations above, we can obtain
the 1- and 2-halo term predictions for the projected mass profile
$\Delta \Sigma(R)$ to compare with the measurement. 
The above
equations also  show that the dimension of the power spectrum
$C_{\Sigma g}(k)$ is in units of $[M_\odot]$; the LRG-shear correlation
is proportional to $k^2 C_{\Sigma g}(k)$ (see Eq.~\ref{eq:dsigma_def}), 
which has  dimensions of
surface mass density in units of $[M_\odot/{\rm Mpc}^2]$.
Given statistical uncertainties of the measurements for a small sample
size of the multiple-LRG systems, we will use the above equations to
constrain the {\em averaged} halo parameters and off-center parameters
for the multiple-LRG systems. 

\begin{figure}
\begin{center}
\includegraphics[width=8.5cm,angle=-90]{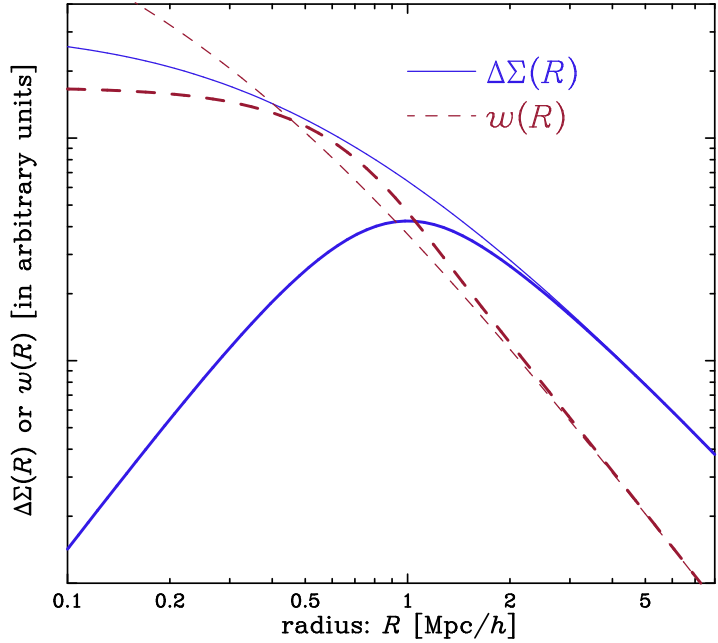}
\caption{The plot demonstrates how the off-centering effect alters
the LRG-galaxy weak lensing profile (solid curves) and the projected
correlation function of LRGs with photometric galaxy positions (dashed)
as a function of the projected radius from the true center or 
LRG positions, which are taken as the ``center'' 
in the correlation calculation for their respective thin and thick
 curves, respectively. 
The amplitude of each profile
 (in the $y$-axis)
is plotted
in arbitrary units.  We assumed
$M_{180b}=1.6\times 10^{14}M_\odot/h$ and $c_{180b}=4.8$ for the halo
mass and concentration parameters, respectively, and employed $R_{\rm
off}=400~{\rm kpc}/h$ for the off-centering parameter, where we assumed that 
all the 
 LRGs are off-centered (see Eq.~\ref{eq:poff}).
The 
weak lensing distortion 
probes the gravitational tidal field around lensing
halos and the cross-correlation of LRGs with galaxies probes the projected
number density profile of galaxies in the host halos.  
The off-centering effect dilutes
the correlation function amplitudes at radii smaller than the
off-centering radius. However, note different features of the off-centering
 effects on the two profiles. For the projected correlation function, 
the off-centering effect
causes an {\em enhancement} in the amplitude
 over a range of the intermediate radii; 
 $w_{\rm
off}(R)>w_{\rm w/o~ off}(R)$. On the other hand, the inequality 
$\Delta\Sigma_{\rm
 off}(R)\le
\Delta\Sigma_{\rm w/o~ off}(R)$ always holds
for the LRG-galaxy lensing profile due to the non-local nature of the
 tidal field. 
\label{fig:off_model} }
\end{center}
\end{figure}
The thin and thick solid curves in Fig.\ref{fig:off_model} show how
the off-centering effect dilutes the lensing profile. Here we only
considered the 1-halo term, which is the dominant term on the small scales we
are interested in. For model parameters, we assumed $M_{180b}=1.6\times
10^{14}~M_\odot/h$ and $c_{180b}=4.8$  for the halo mass and
concentration parameters of an NFW profile, and employed $R_{\rm
off}=400~{\rm kpc}/h$ for the off-center profile parameter (see
Eq.~\ref{eq:poff}); these are typical
values for multiple-LRG systems as we will find below. 
The off-centering effect dilutes the
amplitude of lensing signal; 
$\Delta \Sigma_{\rm off}(R)<\Delta \Sigma_{\rm w/o~ off}(R)$ at 
radii smaller than the off-centering radius, and then
$\Delta \Sigma_{\rm off}(R)=\Delta \Sigma_{\rm w/o~ off}(R)$ at 
larger radii. These features are useful in interpreting the lensing
signals measured from the SDSS DR7 catalog.

When we consider the case that only some fraction of LRGs have
offsets, and the remaining LRGs are central galaxies in their host
halos, we replace the mass profile in the above
equation with 
\begin{equation}
\bar{M}\tilde{u}_{{\rm NFW}}\!\left(k; \bar{M},z_{\rm LRG}
\right) \tilde{p}_{\rm off}(k; \bar{M})
\longrightarrow
\left[
q_{\rm cen}
+ 
(1-q_{\rm cen})
\tilde{p}_{\rm off}(k; \bar{M})
\right]
\bar{M}
\tilde{u}_{{\rm NFW}}\!\left(k; \bar{M},z_{\rm LRG}\right),
\label{eq:qcen_wl}
\end{equation}
where $q_{\rm cen}$ is the fraction of central LRGs in the sample, and
we assumed that all the LRGs reside in halos of the same (or similar)
mass scale $\bar{M}$. In this case, the sub-halo lensing contribution
can be considered as the mean mass of sub-halos hosting the central and
satellite LRGs. For the central LRGs, this enhancement relative to the
NFW lensing of host halos might be considered to model additional lensing
contribution due to baryonic contraction at the halo center. However,
the sub-halo and baryonic contraction contributions are generally
degenerate, unless we know which LRGs in the catalog are
central galaxies. Hence the sub-halo contribution in the model (Eq.~\ref{eq:lens_2h_app})
can be considered as the average of the
two effects. 

\subsection{The projected cross-power spectrum of off-centered LRGs with
 photometric galaxies }
\label{sec:angcor}

The third observable we consider is the projected cross-correlation
function of 
LRGs with positions of photometric-redshift
galaxies, measured as a function of the projected separation as in the
LRG-galaxy weak lensing (see Eq.~\ref{eq:dsigma_def}). 
Even in the presence of photometric redshift errors (photo-$z$
errors), the cross-correlation method is very powerful in a sense that
it can {\em statistically} discriminate, from photo-$z$ outliers, photometric galaxies that
actually cluster with LRGs at the same redshift\footnote{At small
  scales where lensing magnification is important, there might seem to
be some additional clustering with photometric galaxies; this should
also be modeled if those scales are used for the measurements  (unlike in this work).}
\citep{Newman:08}. While including photo-$z$ outliers in the analysis
 dilutes the cross-correlation
amplitude, it does not alter the shape of the correlation
function.

  Similar to the formalism for the LRG-galaxy lensing signal, the
projected cross-correlation function can be expressed in terms of the
power spectrum as
\begin{equation}
w^{\rm cross}_{gg}(R)=
\int\frac{k\rmd k}{2\pi} C^{\rm cross}_{ gg}(k)J_0(kR),
\end{equation}
where the notation ``cross'' is introduced to explicitly mean the
cross-correlation between different populations of galaxies,
spectroscopic LRGs and photometric galaxies, not the auto-correlation
function of the same galaxy population. 
Based on the halo model, the galaxy power
spectrum is given as $C^{\rm cross}_{gg}(k)=C^{{\rm cross}, 1h}_{gg}(k)
+C^{{\rm cross}, 2h}_{gg}(k)$ with the
1- and 2-halo terms being defined as
\begin{eqnarray}
C_{gg}^{{\rm cross}, 1h}(k)&\equiv& 
\int\!\rmd\chi f_{\rm LRG}(\chi)f_{\rm phg}(\chi)
\int\!\rmd M~\frac{\rmd n}{\rmd M}N^{\rm LRG}_{\rm HOD}(M) 
N^{\rm phg}_{\rm HOD}(M) 
\tilde{u}_{{\rm NFW}}\!\left(k; M,z
\right)
\tilde{p}_{\rm off}(k; M),
\nonumber\\
C_{gg}^{{\rm cross}, 2h}(k)&\equiv& 
\int\!\rmd\chi f_{\rm LRG}(\chi)f_{\rm phg}(\chi)
\left[
\int\!\rmd M~\frac{\rmd n}{\rmd M}N^{\rm LRG}_{\rm HOD}(M)b(M)\right]
\left[
\int\!\rmd M'~\frac{\rmd n}{\rmd M'}N^{\rm phg}_{\rm HOD}(M')b(M')\right]
P_m^L\left(k;z\right),
\end{eqnarray}
where $N^{\rm phg}_{\rm HOD}(M)$ is the halo occupation distribution for
photometric galaxies, and we have assumed that the radial profile of
photometric galaxies within the host halos follows an NFW profile for
simplicity (see below for the analysis relaxing the assumption).  
We again note that, in the following results, we will
focus on the cross-correlation at scales up to a few Mpc, so our
 use of the linear power spectrum in the 2-halo term does not cause any
 serious systematic error.  
We will take into account the uncertainty in the 2-halo
 term by treating the bias parameter as a free parameter, as in the
 LRG-galaxy lensing case. 
 Taking into account the photo-$z$ errors, the
redshift selection function $f_{\rm phg}$ is given as
\begin{equation}
f_{\rm phg}(\chi)\equiv \frac{1}{\bar{n}_{\rm phg, {}all}^{\rm 2D}} 
p(z|z_{\rm phz})\frac{\rmd z}{\rmd\chi},
\end{equation}
where $p(z|z_{\rm phz})\Delta z$ is the probability that photometric
galaxies with photo-$z$ of $z_{\rm phz}$ have a true
redshift in the range $[z-\Delta z/2,z+\Delta/2]$. 
The probability
satisfies the following normalization condition: $\int\!\rmd z \,p(z|z_{\rm
phz})=1$. However, generally $p(z|z_{\rm phz})\Delta z\le 1$ around
the LRG 
redshift, so the
factor $p(z|z_{\rm phz})\Delta z$ represents the dilution effect that is
caused by including photo-$z$ outliers in the correlation analysis. 
The
quantity $\bar{n}^{\rm 2D}_{\rm phg, {}all}$ is the mean surface number 
density of all the
photo-$z$ galaxies, defined as $\bar{n}^{\rm 2D}_{\rm phg, {}all}\equiv
\int\!\rmd\chi~ p(z|z_{\rm phz})(\rmd z/\rmd\chi)
\int\!\rmd M(\rmd n/\rmd M)N_{\rm HOD}^{\rm
phg}(M)$.  The selection function $f_{\rm phg}(\chi)$ satisfies the
normalization condition: $\int\!\rmd\chi f_{\rm phg}(\chi)
\int\!\rmd M(\rmd n/\rmd M)N^{\rm phg}_{\rm HOD}(M) =1$.

Assuming a thin redshift slice and narrow halo mass bin of LRGs, we
can simplify the galaxy power spectra as
\begin{eqnarray}
C_{gg}^{1h}(k)\!\!&\simeq & \!\!
\frac{1}{\displaystyle \Delta \chi
\left.\frac{\rmd n}{\rmd M}\right|_{M=\bar{M}}
\Delta M}
\frac{\bar{n}^{\rm 2D}_{\rm phg}(z_{\rm LRG})}{\bar{n}^{\rm 2D}_{\rm phg,{} all}}
\tilde{u}_{{\rm NFW}}\!\left(k; \bar{M},z_{\rm LRG}
\right)
\tilde{p}_{\rm off}(k; \bar{M}), \nonumber \\
C_{gg}^{2h}(k)\!\!&\simeq& \!\!
\frac{1}{\Delta \chi}
\frac{\bar{n}^{\rm 2D}_{\rm phg}(z_{\rm LRG})}{\bar{n}^{\rm 2D}_{\rm phg,{} all}}
\bar{b}_{\rm LRG}\bar{b}_{\rm phg}
P_m^L\left(k; z_{\rm LRG}\right),
\label{eq:wR_app}
\end{eqnarray}
where $\Delta \chi$ is the radial distance width of the LRG
distribution, and $\bar{n}^{\rm 2D}_{\rm phg}(z_{\rm LRG})$ is the
average surface number density of photometric galaxies that indeed lie
in the LRG redshift bin and reside in halos hosting LRGs, 
defined as $\bar{n}^{\rm 2D}_{\rm phg}(z_{\rm
LRG})
\equiv  
p(z_{\rm LRG}|z_{\rm
phz})\left.\rmd z/\rmd\chi\right|_{z_{\rm LRG}}\Delta \chi \rmd n/\rmd
M|_{\bar
M}\Delta M$.
The dimension of the power spectra is $[{\rm
Mpc}]^{2}$. The prefactor $\bar{n}^{\rm 2D}_{\rm phg}(z_{\rm
LRG})/\bar{n}^{\rm 2D}_{\rm phg, {}all}$ gives the fraction of photometric
galaxies with true redshifts that are sufficiently close to the LRG
redshift. Hence, this factor
accounts for the dilution effect due to 
photo-$z$ outliers that we include
in our sample.
Also note that the 1-halo term of
the power spectrum behaves like the Poisson shot noise $1/[\Delta\chi
(\rmd n/\rmd M)\Delta M]$ in the limits of $\tilde{p}_{\rm off}, \tilde{u}_{\rm
NFW}\rightarrow 1$, where $\Delta\chi (\rmd n/\rmd M)\Delta M$ is the mean
surface number density of halos hosting LRGs. 
In the following analysis, we introduce a nuisance parameter to model
the amplitude uncertainty, and will derive constraints on the
off-centering profile of LRGs, marginalizing over the nuisance parameter.

The dashed curves in Fig.\ref{fig:off_model} demonstrate the
off-centering effect on the projected correlation function. The
off-centering effect dilutes the amplitude at small radii, $w_{\rm
off}(R)<w_{\rm w/o~off}(R)$, however it {\em enhances} the amplitude at the
intermediate radii, $w_{\rm off}(R)>w_{\rm w/o~off}(R)$.  Then the two
correlation functions with and without the off-centering effect agree
at the larger radius.
The enhancement in the correlation amplitude is
contrasted to the off-centering effect on the LRG-galaxy lensing profile, which
probes the non-local tidal field around the LRG halos. 
 Here, although we assume
that the photometric galaxies follow an NFW profile for simplicity,
these qualitative features are still present for any arbitrary profile
of the projected density 
field.  These features are useful when making an interpretation of the
measured cross-correlation functions for the SDSS DR7 data as we will
show below.

When further including the mixture of central and satellite LRGs in the
sample, we will use the replacement given in Eq.~(\ref{eq:qcen_wl}) to
model the power spectrum $C_{gg}(k)$. 
Fourier-transforming the
above equations gives the halo model predictions for the projected
cross-correlation $w^{\rm cross}(R)$ to compare with the measurement. 
We will use the above equations to derive the {\em averaged} halo
parameters and off-centering parameters in the following analysis.

It is important to note that while the prediction for the
LRG-photometric galaxy cross-correlation function on small scales
includes a convolution of the radial distribution of satellite galaxies
with the off-centering distribution of our chosen halo centers, the
power of this method is really in a comparison of results with {\em
different} centers for the same halo and photometric galaxy catalogs.
In this case, the radial distribution of satellite galaxies must be the
same for all measurements, and only the off-centering distribution can
cause differences between the measurements.

\section{Measurements and Results}
\label{sec:results}

\subsection{SDSS DR7 LRG Catalog}
\label{sec:sdss_dr7}

Here we describe the data used for the analysis in this paper, all of
which come from the Sloan Digital Sky Survey Data Release 7
catalog (SDSS DR7).  
The SDSS \citep{2000AJ....120.1579Y} imaged roughly $\pi$ steradians
of the sky, and followed up approximately one million of the detected
objects spectroscopically \citep{2001AJ....122.2267E,
  2002AJ....123.2945R,2002AJ....124.1810S}. The imaging was carried
out by drift-scanning the sky in photometric conditions
\citep{2001AJ....122.2129H, 2004AN....325..583I}, in five bands
($ugriz$; \citealt{1996AJ....111.1748F, 2002AJ....123.2121S}) using a
specially-designed wide-field camera
\citep{1998AJ....116.3040G}. These imaging data were used to create
star and galaxy catalogs that we use in this paper.  All of
the data were processed by completely automated pipelines that detect
and measure photometric properties of objects, and astrometrically
calibrate the data \citep{2001ASPC..238..269L,
  2003AJ....125.1559P,2006AN....327..821T}. The SDSS I/II imaging
surveys were completed with a seventh data release
\citep{2009ApJS..182..543A}, though this work will rely as well on an
improved data reduction pipeline that was part of the eighth data
release, from SDSS III \citep{2011ApJS..193...29A}; and an improved
photometric calibration \citep[`ubercalibration',][]{2008ApJ...674.1217P}.

We use the SDSS DR7 LRGs 
in the Northern Galactic Cap
with a contiguous
region, which is made publicly available by
\cite{Kazinetal:10}. 
The sample consists of 96,762 LRGs with absolute magnitude $-23.2<M_g<-21.2$
in the redshift range $0.16<z<0.47$ covering about $1.58 \,({\rm
Gpc}/h)^3$ comoving volume in the concordance $\Lambda$CDM model, with
nearly constant comoving number density for $z<0.36$ and decreasing
number density above that redshift.
To  
recover the halo density field from the LRG distribution, we used the
Counts-in-Cylinders (CiC) techniques developed by
\citet{ReidSpergel:09} to identify groups of $N_{\rm LRG}\ge 2$ LRGs occupying the same
halo. To be more precise, multiple
LRGs are considered as neighbors of the same group 
when the
transverse separation satisfies $\Delta r_\perp\le 0.8h^{-1}$Mpc and the
redshift difference satisfies $\Delta z/(1+z)\le 0.006$ corresponding
to $\delta v_p=1800~$km/s. The radial-direction length of the cylinder is
motivated by the virial motion of galaxies in a massive halo, and was shown to safely
include satellite galaxies in the same halo using mock
catalogs 
\citep{Reidetal:09}.
Then LRGs in different CiC groups are grouped together by a
Friends-of-Friends method (by connecting any of the member galaxies in
the different CiC cylinders).  Table~\ref{tab:lrgs} summarizes the
reconstructed LRG groups. The total number of 
halos inferred from the LRG catalog is  92,046, and about
4.5 per cent are multiple-LRG systems.  Among the multiple-LRG systems, about 90 per cent
are systems with two LRGs ($N_{\rm LRG}= 2$).  We also generate random
catalogs that have the same area coverage and redshift distribution
as the  halos; these are used  to estimate the LRG-galaxy
lensing signal and the cross-correlation with photometric 
galaxies in the imaging data.

Although this CiC method was carefully studied in
\citet{ReidSpergel:09}, it might provide an imperfect catalog of
LRG-host halos; for example, single-LRG systems might be in multiple
LRG-systems, if other LRG candidate(s) 
lack(s) spectra 
due to fiber collisions.  However, given the different goal of this
work, we
do not explore a modification of the CiC method. Rather, we will focus
on a possible residual contamination of off-centered LRGs 
and discuss the impact on the 
 LRG clustering analysis, by using the same method of halo
reconstruction in the previous study. The off-centering effect or the
offset of LRGs from the true center of dark matter halo
is very
difficult to observationally constrain, and we hope our method gives a
way to observationally constrain the off-centering effect.

\begin{table*}
\begin{center}
\begin{tabular}{l||l}
  \hline\hline Number of LRG(s) ($N_{\rm LRG}$) in each FoF group &
Number of LRG FoF groups (fraction) \\ \hline 1 & 87889 (95.5 per cent)
\\ 2 & 3713 \\ 3 & 358 \\ 4 & 65 \\ 5 & 14 \\ 6 & 6 \\ 7 & 1 \\ \hline
\hline Total & 92046 (100 per cent) \\ \hline
\end{tabular}
\end{center}
\caption{The number of different LRG groups after the CiC-FoF based
 group finder of the SDSS DR7 LRG distribution. ``$N_{\rm LRG}=1$''
 means single LRG systems, where each region has a single LRG in the
 cylinder region in angular and redshift space. ``$N_{\rm LRG}\ge
 2$'' systems are multiple-LRG systems, where each region contains
 $N_{\rm LRG}$ LRGs in the cylinder region.
\label{tab:lrgs}
}
\end{table*}

\begin{figure}
\begin{center}
\includegraphics[width=8.6cm]{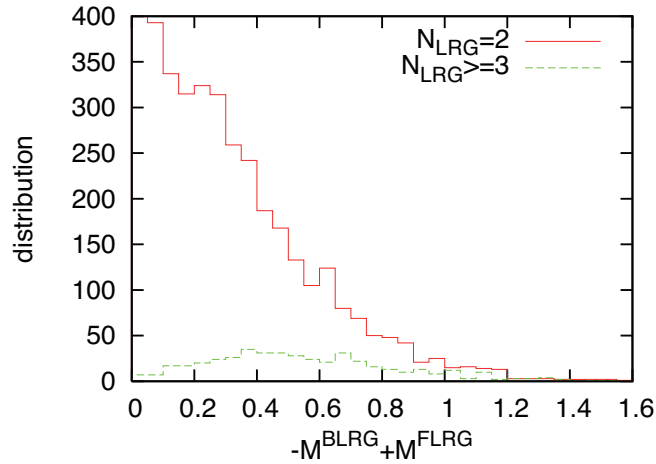}
\caption{The histogram of magnitude differences between
 the brightest and faintest LRGs (BLRG and FLRG) in  multiple-LRG
 systems (see Table~\ref{tab:lrgs}). The solid-line histogram is for the systems
 having 2 LRGs inside, while the dashed line is for the systems with
 $N_{\rm LRG}\ge 3$. 
\label{fig:LRGs_dm}
}
\end{center}
\end{figure}
To compute the power spectrum and the cross-correlations from the
LRG-inferred halos, 
we must 
define the halo center for each LRG system. (1) For single LRG systems
with $N_{\rm LRG}=1$, we use the redshift and angular position of LRG as
the halo center proxy. (2) For multiple-LRG systems with $N_{\rm LRG}\ge 2$,
we use
the following three halo center proxies in order to compare the measurements: 
\begin{itemize}
 \item {\em BLRG}: the brightest LRG (BLRG) in each multiple-LRG group as
      the halo center. The BLRG is expected to 
reside in the most massive
      sub-halo and therefore be closer to the underlying true center
      due to dynamical friction theory.
\item {\em FLRG}: the faintest LRG (FLRG) as the halo
      center; it can be considered as the extreme counterpart to
      BLRG. 
\item {\em Mean}: the mean position of LRGs in their redshift and
      angular positions. Note that we did not use any luminosity or
      other weighting to estimate the mean position. This 
is the halo-center proxy
      used in \cite{Reidetal:10}. 
\end{itemize}

Fig.\ref{fig:LRGs_dm} shows the distribution of magnitude difference
between the brightest and faintest LRGs (BLRG and FLRG) in the 
multiple-LRG systems. The magnitude difference is typically within 1 mag. 
Most of the 2-LRG systems have a magnitude difference that is  less
than a few tenths of a magnitude, implying that the BLRG and FLRG are
not very 
different, and the FLRG may be closer to the true
center than the BLRG in some halos, as we will address below.

\subsection{LRG power spectrum}

\begin{figure}
\begin{center}
\includegraphics[width=8.6cm]{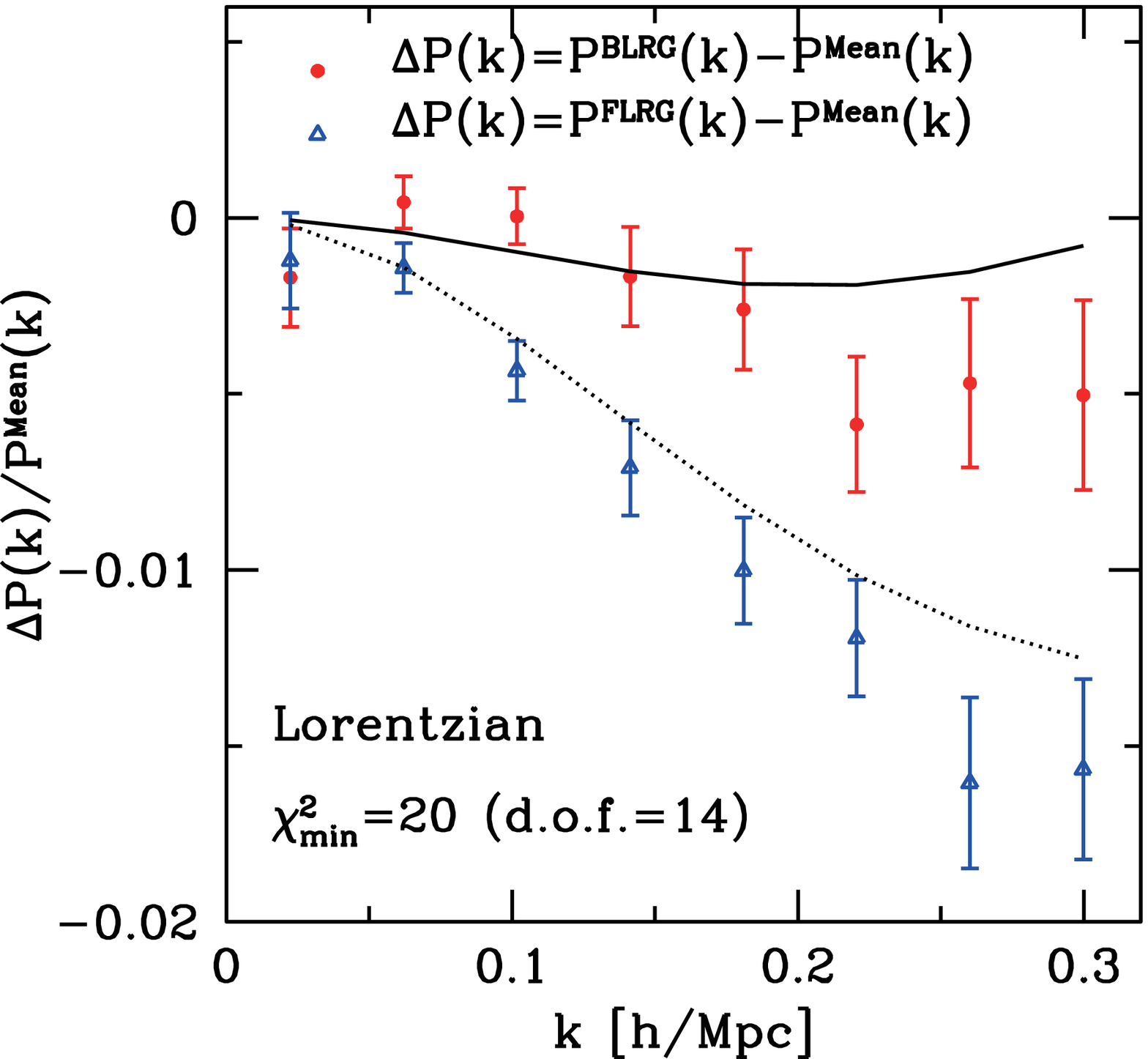}
\includegraphics[width=8.6cm]{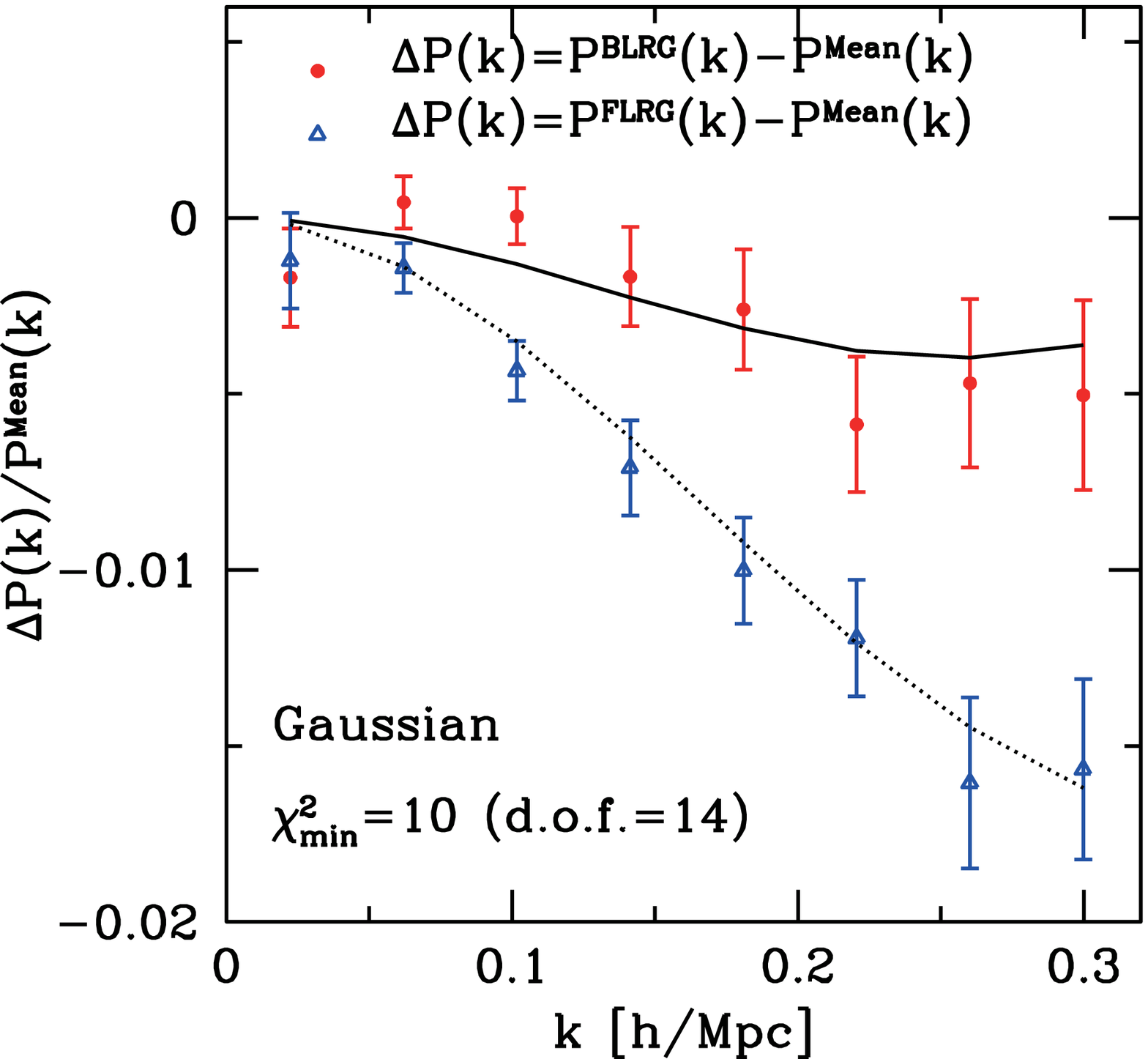} 
\caption{The angle-averaged redshift-space power spectra for the
 LRG-inferred halos, measured using the different halo center proxies
 for  the
 multiple-LRG systems: the brightest LRG position (BLRG), the faintest
 LRG position (FLRG) and the mean position (Mean) of
 LRG positions in angular and redshift space (see Section~\ref{sec:sdss_dr7}
 for the details).  The circle and triangle symbols show
 the fractional differences of the halo
 power spectra of the BLRG or FLRG 
 centers, respectively, relative to the spectrum of the Mean center.
Since we used the same halos and 
 power spectrum measurement methods, 
the difference arises solely from the different halo centers of the
 multiple-LRG systems. 
The FLRG power spectrum shows a larger
suppression compared to the other two, which indicates that the FLRGs
 have the largest offsets 
and thus largest internal velocity dispersion 
in the multiple-LRG systems. The
BLRG power spectrum still shows some suppression compared to that for the Mean, 
which indicates a residual off-centered effect of BLRGs. 
The solid and dashed curves in each panel show the best-fit FoG model
 assuming the Lorentzian ({\em left panel}) and Gaussian ({\em right})
 functional forms. Note that the measurement points are the same in both
 panels. 
}  \label{fig:fog}
\end{center}
\end{figure}

We first use the angle-averaged, redshift-space power spectrum of
LRG-inferred halos
to study the effect of off-centered LRGs.  We employed the method of
\cite{Reidetal:10} to measure the power spectrum from the reconstructed
halo density field, as briefly summarized below.

The power spectrum measurement 
was done using the Fourier
 decomposition
 method
developed in \cite{Feldmanetal:94}. We used $1024^3$ grids in a cubic box
with side length 
$2.6~{\rm Gpc}/h$,
which covers the entire region of
the LRG distribution in redshift space. The density fluctuation field of
halos on a grid is estimated by placing the halos onto the grid and
subtracting an unclustered ``random'' catalog, which matches the halo
selection. The random catalog is given from \cite{Kazinetal:10}, where
the variations in angular and radial selections and the effect of fiber
collisions are taken into account. In doing the density assignment from
 the LRG-halos or random catalogs, we used the weighting method
 in \cite{Percivaletal:04}
 using a luminosity-dependent bias model.
However, note that we used the luminosity of BLRG in each halo to
compute the weighting, and used the {\em same} weight for all the three
BLRG, FLRG and Mean centers in the multiple-LRG systems.  Thus we used
the {\em same} random catalogs and the {\em same} luminosity weights for
all the three BLRG-, FLRG- and Mean-center catalogs in order to avoid
any unwanted effect on the measured power spectra. We also
note that, in the following results, we focus 
on relative {\em differences}
 between the measured power spectra and
cross-correlation functions using the different centers, 
where the differences should not be sensitive to
any details of the  measurement methods.

To estimate the error bars or measurement uncertainties, 
we divide the LRG sample
into 100 equal-area, nearly contiguous regions on the sky, and carry
out the measurements of the LRG-inferred power spectra from each of 
 those 100 regions \citep[][hereafter M12]{Mandelbaumetal:12}; 
for some description of the limitations of bootstrap and
jackknife resampling, see, e.g., \cite{2007A&A...464..399H} and
\cite{2004MNRAS.353..529H}.  Then we estimate the covariance matrix
from the 100 power spectra. 
The jackknife resampling assumes
independence of the regions, and thus cannot be used to obtain the error bars
for scales comparable to the size of the regions (which are far larger than the maximum scales used for our fits). 

Fig.\ref{fig:fog} shows the measured power spectra of halos
reconstructed from the SDSS LRG catalog. Plotted here is the fractional 
difference
of the power spectra, measured by using either BLRG- or FLRG-halo centers,
relative to the power spectrum calculated using the Mean-center in the multiple-LRG systems. The
error bars at each $k$ bin are estimated from the jackknife method we
described above. Note that, in this analysis, we changed the
redshift- and angular positions
 only for 4.5 per cent of the halos
(the other 95.5 per cent are exactly the same since they include only
single-LRG systems). 
The figure clearly shows that the power spectra calculated using both 
the BLRG and FLRG centers show smaller amplitudes with
increasing $k$
than the power spectrum calculated using 
the Mean center, implying that the BLRG and FLRG results 
are more affected by the FoG effect, leading to 
a stronger suppression in the power than the
Mean spectrum. 
The
significance of the differences is $\Delta \chi^2=69, 120$
and $138$ for the BLRG, FLRG and the combined BLRG and FLRG difference,
respectively, corresponding to about 8.2, 10.9, and 11.7$\sigma$
detections of the FoG effects, respectively. In these estimates, we
included the correlations between the error bars at different $k$ bins
(which are typically around $0.1$, but occasionally as large as
$0.5$), 
and the (significant) correlation between the BLRG and FLRG spectra using the
full jackknife covariance matrix.
The FLRG spectrum has more suppressed power, which is 
expected since 
fainter galaxies are more likely to be satellites, 
but it is more surprising that the
BLRG  power spectrum also shows the FoG suppression. This suggests that 
the BLRGs also have an off-centered distribution in the host halos, or
some of the BLRGs are satellite galaxies. 
On the other hand,  the power spectrum of the Mean center 
turns out to be least
affected by the FoG effect, which is counter-intuitive, because there is
no LRG at the inferred position.  
However, we note that most of the multiple-LRG
systems have $N_{\rm LRG}=2$ and a small magnitude
difference between 
the two LRGs. Hence,
some FLRGs, rather than the BLRGs, may be central galaxies in some of
 the host halos. 

Following the method developed in \cite{Hikageetal:12}, we can make a
model interpretation of the measured power spectra for the different
LRG-inferred halo centers. The redshift-space power spectrum we measure
arises from auto- and cross-correlations of/between the single- and
multiple-LRG halos, 
where the fraction of multiple-LRG halos among all the
 LRG-inferred halos is only 4.5 per cent, which we
hereafter refer
as $f_{\rm
M-LRG}\equiv 0.045$. 
We assume that only some fraction of the chosen LRGs are central
galaxies in the single- or multiple-LRG systems, and represent these
fractions as $q_{\rm
cen}^{\rm S-LRG}$ or $q_{\rm cen}^{\rm M-LRG}$ (to be more precise,
we will consider
$q_{\rm cen}^{\rm BLRG}$ or $q_{\rm cen}^{\rm FLRG}$ for the multiple-LRG
halos, because we chose the galaxies for the halo center proxies). The
remaining galaxies, given by $(1-q_{\rm cen})$, are satellites that
cause the FoG effect. 
Including the parameter
$q_{\rm cen}$ gives a better agreement with measurements of the
LRG-galaxy lensing and the projected cross-correlation function of
LRG-halos and photo-$z$ galaxies, to be described in subsequent
sections. 
Thus, we can model the redshift-space power spectrum of
the LRG-inferred halos as
\begin{eqnarray}
P_{s}(k,\mu)&=& \left[1+\beta\mu^2\right]^2
\left[
\bar{b}_{\rm S-LRG}
(1-f_{\rm M-LRG})
\left\{
q^{\rm S-LRG}_{\rm cen}+(1-q_{\rm cen}^{\rm S-LRG})\sqrt{F_{\rm S-LRG}(k,\mu)}
\right\}\right.\nonumber\\
&&\hspace{6em}\left. + \bar{b}_{\rm M-LRG}
f_{{\rm M-LRG}}
\left\{
q_{\rm cen}^{\rm M-LRG}+(1-q_{\rm cen}^{\rm M-LRG})\sqrt{F_{\rm M-LRG}(k,\mu)}
\right\}
\right]^2P^{\rm NL}_m(k),
\label{eq:ps_s_lrg1}
\end{eqnarray}
where $\bar{b}_{\rm S-LRG}$ and $\bar{b}_{\rm M-LRG}$ are the average
bias parameters for the single- and multiple-LRG host halos,
respectively; $q_{\rm cen}^{\rm S-LRG}$ is the fraction of central LRGs
in the single-LRG systems;  
$q_{\rm cen}^{\rm M-LRG}$ is the fraction of either BLRGs or FLRGs to be
centrals  in the
multiple-LRG systems.  
We have assumed
that the redshift-space distortion effect due to the peculiar velocity
field of halos is given by the Kaiser formula \citep{Kaiser:87}, given
by the function $[1+\beta\mu]^2$, and $\beta\equiv (1/b)\rmd\ln
D/\rmd\ln a$ ($D$ is the linear growth rate). $P^{\rm NL}_{m}(k)$ is the
real-space, nonlinear power spectrum of dark matter, and $F(k,\mu)$ is the
velocity function to model the average velocity distribution of
off-centered (satellite) LRGs within the host halos (see below). 
The coefficients $(1-f_{\rm M-LRG})$ and $f_{\rm M-LRG}$ are the
fractions of the single- and multiple-LRG halos relative  to all the
LRG-host halos, respectively. 
In the above equation, we assumed that the central LRG
rests at the center of the host halo
and therefore has no FoG effect.  Thus, different populations of
LRG-host halos (the single- and multiple-LRG systems in our case) cause
FoG effects on the redshift-space power spectrum in different ways,
varying with their fractions, the halo bias parameters and the fractions
of satellite LRGs in each population. To compute the bias parameters, for
simplicity we employ the halo bias using the average halo masses
measured from the LRG-galaxy weak lensing for the assumed fiducial
cosmology: $\bar{b}_{\rm S-LRG}=2.12$ and $\bar{b}_{\rm M-LRG}=3.26$
for the average masses of the host halos, $\bar{M}_{\rm
S-LRG}=0.4\times 10^{14}~M_\odot/h$ and $\bar{M}_{\rm M-LRG}=1.6\times
10^{14}~M_\odot/h$, respectively, estimated using the bias model in
\cite{ShethTormen:99}\footnote{While we have used a
  theoretical bias vs. halo mass relation, it is worth noting that the
  large-scale biases for LRG samples that together compose the LRG sample in this
  work were found (via direct-fitting to large-scale clustering and
  lensing measurements) to be $2.07\pm 0.05$ and $2.26\pm 0.06$
  \citep{Mandelbaumetal:12}.  Thus we expect that the average
  of $\bar{b}_{\rm S-LRG}$ and $\bar{b}_{\rm M-LRG}$ (weighted by the
  relative fractions of single- and multiple-LRG systems) to lie in
  between those two measurements, which is indeed the case given the
  predominance of single-LRG systems.  This confirms that use of these
  theoretical predictions is consistent with the 
  data.}.  
For the
velocity function $F$, we assume the following functional forms that are
often used in the literature \citep{Hamilton:98}:
\begin{eqnarray}
F(k,\mu)=
\left\{
\begin{array}{ll}
{\displaystyle 
\frac{1}{[1+(k\mu\sigma_{v,{\rm off}}/aH(z))^2] }},
& \mbox{(Lorentzian FoG effect)}\vspace{0.5em}\\
{\displaystyle 
\exp[-(k\mu\sigma_{v,{\rm off}}/aH(z))^2] },
& \mbox{(Gaussian FoG effect)}
\end{array}
\right.
\label{eq:fog}
\end{eqnarray}
where $\sigma_{v,{\rm off}}$ is the typical velocity dispersion of
off-centered LRGs.  Note that ``$\sqrt{~}$'' of $\sqrt{F}$ in the FoG
effect of Eq.~(\ref{eq:ps_s_lrg1}) was introduced because the above
functions are conventionally used to model the FoG effect on
correlations between {\em two} satellite galaxies. Thus
Eq.~(\ref{eq:ps_s_lrg1}) is different from the form  used in the
literature, and predicts a non-trivial scale dependence as a function of
wavenumber; e.g., at the large-$k$ limit, the FoG function
asymptotically approaches a constant value, $ [\bar{b}_{\rm S-LRG}
(1-f_{\rm
M-LRG})q_{\rm cen}^{\rm S-LRG}+\bar{b}_{\rm M-LRG}f_{{\rm M-LRG}}
q_{\rm cen}^{\rm M-LRG}]^2 $, 
while the standard FoG function $F\rightarrow 0$.

The fractional power spectra shown in Fig.\ref{fig:fog} are with
respect to the power spectrum for the ``Mean'' centers in the
multiple-LRG systems. 
Hence we need to model the power 
spectrum for the Mean center based on our method. 
Fig.\ref{fig:fog} implies that the Mean centers, the
mean position of BLRG and FLRG in multiple-LRG systems, are statistically closer to the
true centers, such that the power spectrum is least affected by the FoG
effect.  
Physically, this is very difficult to imagine, since we expect that 
the Mean center, which has no corresponding galaxy at the position,
should be offset from the halo center to some degree. 
To consider the offset probability of the Mean centers, we need to
include three cases: (1) within the host halo, BLRG is a central galaxy,
while FLRG is a satellite galaxy, (2) 
 FLRG is central, while BLRG is satellite, and (3) both BLRG and FLRG are
satellites in their host halo. 
Among all the multiple-LRG systems, 
the probabilities of these 3 cases are given by
$q_{\rm cen}^{\rm BLRG}$, $q_{\rm cen}^{\rm FLRG}$ and $1-q_{\rm
cen}^{\rm BLRG}-q_{\rm cen}^{\rm FLRG}$, respectively.
With these in mind, the velocity dispersion of the Mean center
can be estimated as
\begin{eqnarray}
\left(\sigma_{v, {\rm off}}^{{\rm Mean}}{}\right)^2
&=&
q_{\rm cen}^{\rm BLRG}
\frac{\left(\sigma_{v,{\rm off}}^{\rm
				  FLRG}\right)^2}{4}
+
q_{\rm cen}^{\rm FLRG}
\frac{\left(\sigma_{v,{\rm off}}^{\rm
				  BLRG}\right)^2}{4}
+
\left(
1-q_{\rm cen}^{\rm BLRG}-q_{\rm cen}^{\rm FLRG}
\right)
\frac{
\left(\sigma_{v,{\rm off}}^{\rm	  BLRG}\right)^2
+
\left(\sigma_{v,{\rm off}}^{\rm	  FLRG}\right)^2
}{4}
\nonumber\\
&=&\frac{(1-q_{\rm cen}^{\rm BLRG})
\left(\sigma_{v,{\rm off}}^{\rm	  BLRG}\right)^2
+
(1-q_{\rm cen}^{\rm FLRG})
\left(\sigma_{v,{\rm off}}^{\rm	  FLRG}\right)^2}{4},
\end{eqnarray}
where the three terms in the first line correspond to the three cases
mentioned above, and 
we have assumed no correlation between the distributions of BLRG
and FLRG in each host halo as well as no correlation between the
distributions of LRGs in different halos. We also 
assumed that BLRG and FLRG
determine 
the mean position of LRGs in multiple-LRG systems, which may host more
than two LRGs ($N_{\rm LRG}\ge 3$).  
Or equivalently we assumed that the
Mean centers are determined mainly by the multiple-LRG systems with 2
LRGs, because the systems with 2 LRGs are a dominant population of the
multiple-LRG systems (see Table~\ref{tab:lrgs}).
Thus we expect that the Mean centers 
always have offsets from the true center. 
The factor of 4 in the denominator of each term is from the fact that we
computed the dispersion of the off-centering amount of the Mean centers.
For instance, in the first case (the first term in the above equation),
suppose the line-of-sight displacement vector for the Mean center with
respect to the true center 
in each system is defined
 as $\vec{\delta r}_{\parallel,{\rm Mean}}=\vec{\delta r}_{\parallel, {\rm
FLRG}}/2$ as the BLRG is at the halo center ($\vec{\delta r}_{\parallel, {\rm
BLRG}}=0$). Then the average $\ave{\vec{\delta r}_{\parallel, {\rm Mean}}}=0$ 
and the dispersion 
$(\sigma^{\rm Mean}_{{\rm off}})^2=\ave{(\vec{\delta r}_{\parallel, {\rm
Mean}})^2}=(\sigma_{\rm off}^{\rm FLRG})^2/4$, yielding the
factor of 4. 

Hence, the fractional power spectrum shown in Fig.\ref{fig:fog} can be
modeled based on our method, e.g. for the BLRG center, to first order in
$f_{\rm M-LRG}$: 
\begin{equation}
\frac{\bar{P}_s^{\rm BLRG}(k)}{\bar{P}_s^{\rm Mean}(k)}-1\simeq 
2f_{\rm M-LRG}
\left(
\frac{\bar{b}_{\rm M-LRG}}{\bar{b}_{\rm S-LRG}}
\right)
\frac{\displaystyle 
\int_{-1}^{1}\!\frac{d\mu}{2}
(1+\beta\mu^2)^2\left[q_{\rm cen}^{\rm BLRG}+(1-q_{\rm cen}^{\rm BLRG})
\sqrt{F(k,\mu; \sigma_{v,{\rm off}}^{\rm BLRG})}
-\sqrt{F(k,\mu; \sigma_{v,{\rm off}}^{\rm Mean})}
\right]
}
{\displaystyle \int_{-1}^{1}\!\frac{d\mu}{2}(1+\beta\mu^2)^2}.
\end{equation}
Thus the quantities such as $P^{\rm NL}(k)$ (or $\beta$
to a good approximation) cancel out in the fractional difference, and 
the ratio of bias parameters $(\bar{b}_{\rm
M-LRG}/\bar{b}_{\rm S-LRG})
$ is relevant. 
In the above equation, we also assumed that 
the FoG effect for the single-LRG systems is
negligible,
because the halos have
 a higher fraction of the central LRGs ($q_{\rm cen}^{\rm S-LRG}\simeq
 0.8$) and a smaller velocity dispersion due to the smaller halo
 mass as we will show in the following. 
Since $\sigma_{v,{\rm off}}^{\rm Mean}$ is given in terms of $\sigma_{v,{\rm
off}}^{\rm BLRG}$ and $\sigma_{v,{\rm off}}^{\rm FLRG}$ (and
we have fixed the background cosmological model to compute the
linear Kaiser factor $\beta$), 
the relevant
fitting parameters are 4: 
 $q^{\rm BLRG}_{\rm cen}$,  $q^{\rm FLRG}_{\rm cen}$, 
$\sigma_{v,{\rm off}}^{\rm BLRG}$,
and $\sigma_{v,{\rm off}}^{\rm FLRG}$, respectively. 

The solid and dotted 
curves in the left- and right-panels of Fig.\ref{fig:fog}
show 
the best-fit FoG effect
for the Lorentzian or  
Gaussian FoG models, respectively. 
In this fitting, for simplicity, we assumed that the 
fraction of 
central BLRGs or central FLRGs 
is $q_{\rm cen}^{{\rm BLRG}}=0.54$ or
$q_{\rm cen}^{{\rm FLRG}}=0.32$, respectively, as implied from the 
measurements of the LRG-shear lensing and the LRG-photo-$z$
galaxy cross-correlation in the subsequent sections. 
We find that 
the best-fit velocity dispersions for the satellite BLRGs and FLRGs are
$\sigma_{v,{\rm off}}^{\rm BLRG}=519\pm 48$~km/s and $\sigma_{v,{\rm
off}}^{\rm FLRG}=561\pm 37$~km/s 
for the Lorentzian FoG effect, 
and
$\sigma_{v,{\rm off}}^{\rm BLRG}=498\pm 41$~km/s and $\sigma_{v,{\rm
off}}^{\rm FLRG}=512\pm 32$~km/s 
for the Gaussian FoG effect, 
respectively. 
For the Mean center, 
 $\sigma^{\rm Mean}_{v,{\rm off}}\simeq 291\pm 20$~km/s 
and $\sigma^{\rm Mean}_{v,{\rm off}}\simeq 270\pm 17$~km/s 
for the Lorentzian and Gaussian FoG models, respectively. 
Fig.\ref{fig:fog} shows that the Gaussian FoG model gives a 
better fit to the scale dependence of the FoG effect measured from the
SDSS LRG catalog than the Lorentzian model. 
The minimum $\chi^2$ values inferred are 
for the best-fit models obtained by jointly fitting the model
predictions to  
 the BLRG and FLRG power spectra
taking into account  the
cross-covariance.  The $\chi^2$ has  fourteen degrees of freedom: sixteen data
points minus two model parameters (two $\sigma_{v, {\rm off}}$ parameters).
These results are confirmed by the results using the 
LRG-galaxy weak lensing and the LRG-photo-$z$ galaxy cross-correlation.

\subsection{LRG-galaxy  weak lensing}
\label{sec:weaklens}

As the second observable, 
we use the cross-correlation of
LRG-inferred centers with shapes of background galaxies -- the so-called galaxy-galaxy
lensing using the LRGs as lenses, which was measured 
in the same way as in M12.
In brief, this measurement uses a
catalog of 1.2 background galaxies per arcmin$^2$ with measured
shapes from the re-Gaussianization method \citep{2003MNRAS.343..459H}
and photometric redshifts from Zurich Extragalactic Bayesian Redshift
Analyzer \citep[ZEBRA,][]{2006MNRAS.372..565F}; the catalog was described
and thorough studies of systematic errors were carried out in several
recent papers 
\citep[M12, and also ][]{2011arXiv1110.4107R,2012MNRAS.420.1518M,2012MNRAS.420.3240N}.

The lensing measurement is done as follows. We first 
identify background galaxies around
each LRG that have 
photometric redshift larger than the LRG spectroscopic redshift. 
The weights are assigned to each LRG-background galaxy pair 
using optimal weighting according 
to the error on the source shape measurement via
\begin{equation}
w_{ls} = \frac{1}{\Sigma_{\rm crit}^{2}(\sigma_s^2 + \sigma_{SN}^2)},
\end{equation}
where $\sigma_s^2$ is the shape measurement error due to pixel
  noise, and $\sigma_{SN}^2$ is the RMS intrinsic
ellipticity (both quantities are per component, rather than
  total; the latter is fixed
to $0.365$ in M12.
$\Sigma_{\rm
crit}$ is the critical surface mass density defined by $\Sigma_{\rm
crit}^{-1}(z_l,z_s)\equiv 4\pi G c^{-2}D_{ls}D_l(1+z_l)^2/D_s$,
where $D_l$ and $D_s$ are the angular diameter distances
to lens (LRG) and source and $D_{ls}$ is the distance between them, and
we used the best-fit photo-$z$ of each source to compute $D_s$; this
gives rise to a bias in the signals that can be easily corrected using
the method from \cite{2012MNRAS.420.3240N}. The
factor $(1+z_l)^2$ arises due to our use of comoving coordinates. 
The factor $\Sigma_{\rm crit}^{-2}$ downweights pairs that are close in
redshift and therefore are inefficient in weak lensing.  

The projected mass density  in each radial 
bin can be computed via
a summation over lens-source pairs ``$ls$'' and random lens-source pairs
``$rs$'':
\begin{equation}
\Delta\Sigma(R) = \frac{\sum_{ls} w_{ls} e_t^{(ls)}\!(R) 
\Sigma_{{\rm crit}}(z_l,z_s)}{2 {\cal
    R}\sum_{rs} w_{rs}},
\end{equation}
where $e_t$ is the tangential ellipticity component of source
galaxy with respect to the lens position, 
the factor of $2{\cal R}$ is necessary to convert to tangential shear $\gamma_t$ due to our definition of
ellipticity, and $R$ is the comoving projected radius from lens, which
is estimated from the separation angle between lens (LRG) and source on
the sky as $R=\chi_{\rm LRG}\theta$. 
The division by $\sum w_{rs}$ is necessary to account for the fact
that some of our `sources' are physically associated with the lens,
and therefore not lensed by it.  Finally, we subtract off a similar
signal measured around random lenses, to subtract off any coherent
systematic shear contributions \citep{2005MNRAS.361.1287M}; this
signal is statistically consistent with zero for all scales used in
this work.

To calculate the error bars, we used the jackknife resampling method, as
for the power spectrum measurement. 
The maximum
scale used for the fits in the lensing analysis is 
a few Mpc, within a factor of two of the virial radius of massive
halos; this 
corresponds to 
several arcminutes 
at the typical LRG redshift, which is
$<0.01$
times the typical size of each jackknife resampled region.  Thus, the
jackknife method is
a reasonable approach to getting the covariance matrix for the projected
mass profile. 

\begin{figure}
\begin{center}
\includegraphics[width=8.6cm]{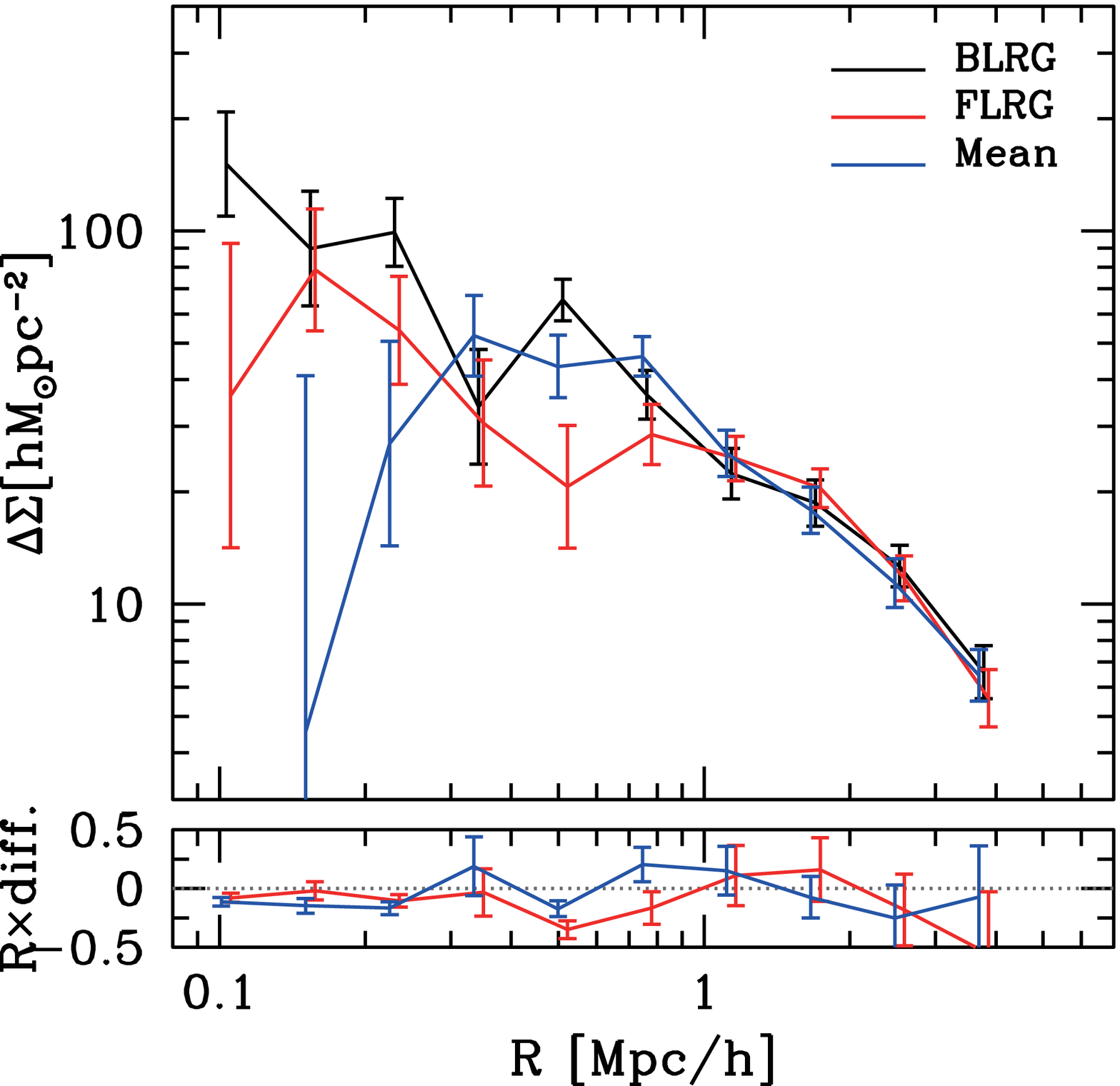}
\includegraphics[width=8.6cm]{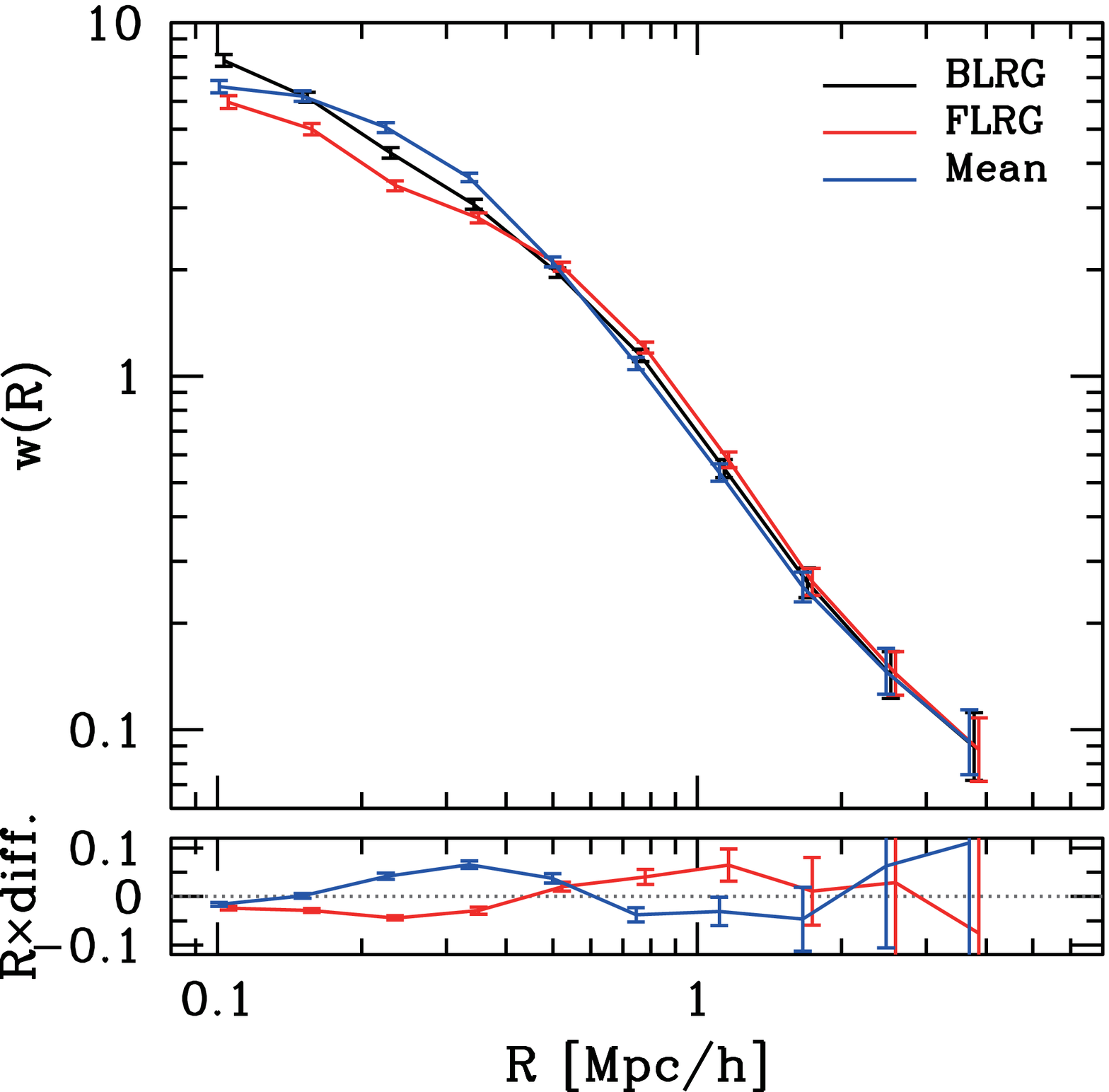}
\caption{{\em Left panel}: 
The cross-correlation of LRG-inferred halos
 with shapes of background, photometric galaxies -- 
the LRG-galaxy weak
 lensing, measured for the multiple-LRG systems using the BLRG, FLRG or
 Mean halo center proxies, respectively. 
The
 background galaxy sample consists of photometric galaxies with 
 photometric redshifts larger than
 the LRG redshift (see text for the details). In these lensing
 measurements, we used the exactly same lens-source pairs and random
 point catalogs.  The three lensing signals at radii $\ga 1~$Mpc/$h$ are
 similar, but the signals at smaller radii differ from each
 other. The lower panel shows the fractional difference of the
profile relative to BLRG center, multiplied by $R$ for illustrative
 purpose:
(FLRG/BLRG$-1)\times R$ or 
(Mean/BLRG$-1)\times R$. The error bars for the ratio properly take into
 account the cross-covariance of the different profiles, because these
 different-center measurements are from the same multiple-LRG systems
 as well as from the same population of background galaxies and therefore the
 errors of the different profiles are highly correlated with each other
 at each radial bin.
{\em Right panel}: Likewise, the projected cross-correlation
 function of the LRG-inferred halos with photometric galaxies for the
 multiple-LRG systems, using the three different centers. The
 photometric galaxy sample is taken from the photometric galaxies
for which the
 photometric redshift is consistent with the spectroscopic redshift of each LRG
 halo within the photo-$z$ errors.  The measured amplitudes at scales
 $R\simgt 0.2~$Mpc$/h$ show increasing off-centering effects in the order
 of the Mean, BLRG and FLRG centers as in the FoG results in
 Fig.\ref{fig:fog} (see text for details). The BLRG correlation
 shows an excess in the power at the small radii, $R\simlt 200$kpc/$h$. 
These cross-correlations of different centers can be interpreted by
 using  models with a mixture of centered and off-centered
 (satellite) LRGs in the host halos 
(see text for discussion). 
 \label{fig:wl-wR_obs}}
\end{center}
\end{figure}
The left panel of Fig.\ref{fig:wl-wR_obs} shows the  LRG-shear
cross-correlation measured for the multiple-LRG systems, using
 the different halo 
centers (Mean, BLRG and FLRG). 
We used the same pairs of lens and sources, and therefore the
difference in the measured correlations arises from 
 the different centers. We are plotting only the
radii relevant for the virial region of massive halos, up to a few
Mpc. The three mass profiles agree with each other at radii $\ga
1$~Mpc/$h$, reflecting that the three centers are in the same
halos. However, the mass profiles differ at smaller radii.  The
intermediate scales around $0.5$~Mpc/$h$ show that the lensing signals
for the BLRG and Mean centers are higher than the FLRG signal,
consistent with our interpretation of the 
FoG measurements in Fig.\ref{fig:fog}.  At the
smaller scales, the mass profile for the Mean center shows a decreasing
power with decreasing radii, which is a clear signature of the dilution
effect due to the off-centering effect and the lack of a galaxy at that position. On the other hand, the mass
profiles of the BLRG and FLRG centers do not show this dilution
  at the smallest scales. 

\begin{figure*}
\begin{center}
\includegraphics[width=17.5cm]{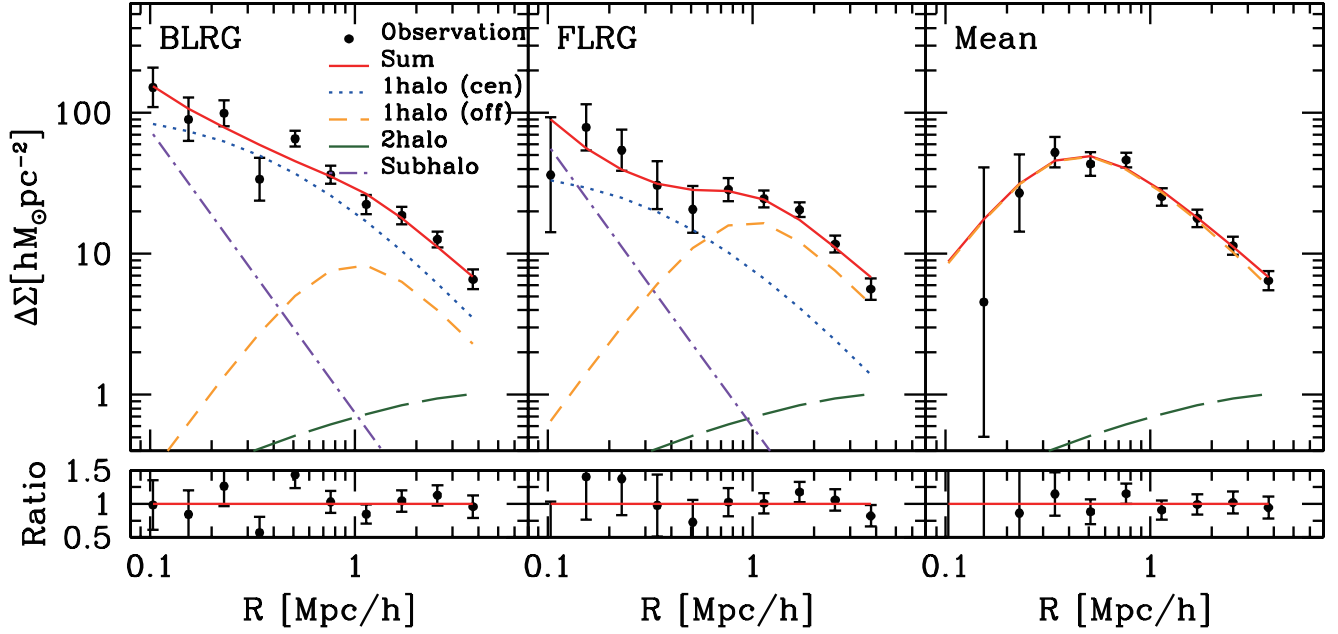}
\caption{The best-fit halo model predictions, developed in
 Section~\ref{sec:wl}, to the measured LRG-galaxy lensing signals, for the 
the BLRG ({\em left panel}), FLRG ({\em middle}) and Mean ({\em right})
centers, respectively. The lower panel in each plot shows the
ratio of  the best-fit model to the measured profile. 
The data with error bars are the same as in the
 left panel of Fig.\ref{fig:wl-wR_obs}. We considered the lensing
 contributions from the host halos of
the multiple-LRG systems and from the sub-halo hosting the BLRG or FLRG
 at small scales. For the host halo contribution, we considered 3
 contributions: the 2-halo term (dashed curve) and the 1-halo terms with
 and without the off-centering effect, indicated by the long- and
 short-dashed curves, respectively. For the BLRG- and
 FLRG-centers, we assumed that some fraction of the BLRGs or FLRGs are
 central galaxies, modeled by $q_{\rm cen}$ (see Eq.~\ref{eq:paras_wl}),
 while the remaining LRGs are off-centered (satellite) LRGs. On the
 other hand, we assumed that all the Mean center have offsets from the
 true center in each host halo. The dot-dashed curve is the sub-halo lensing
 contribution for which we assumed a point mass. The top solid curve in
 each panel is the total lensing signal including all contributing terms. Note that the mass and
 concentration parameters of the host halos are the same in the
 three panels, since the same set of halos was used in each case (we simultaneously fit the model to the
 three measurements).  
}
\label{fig:dsigma}
\end{center}
\end{figure*}
In Fig.\ref{fig:dsigma}, we show the results of fitting the halo model
prediction from Section~\ref{sec:wl} to the measured lensing profiles for the
three centers in the left panel of Fig.\ref{fig:wl-wR_obs}. The fit includes 10 model parameters
\begin{equation}
p_{\alpha}=(\bar{M}_{180b}, \bar{c}_{180b}, q_{\rm cen}^{\rm BLRG}, R_{\rm
 off}^{\rm BLRG}, m_{\rm sh}^{\rm BLRG}, q_{\rm cen}^{\rm FLRG}, R_{\rm
 off}^{\rm FLRG}, m_{\rm sh}^{\rm FLRG},
R_{\rm off}^{\rm Mean},\bar{b}),
\label{eq:paras_wl}
\end{equation}
where $\bar{M}_{180b}$ and $\bar{c}_{180b}$ are the average mass and
concentration parameter of NFW profile  for host halos of the
multiple-LRG systems, $q_{\rm cen}$ is the fraction of the BLRGs or FLRGs 
that are central galaxies in their host halos, $R_{\rm off}$ is the
typical 
off-centering radius (see Eq.~\ref{eq:poff}),  $m_{\rm sh}$ is the
sub-halo mass hosting the BLRG or FLRG, and $\bar{b}$ is the bias
parameter for the 2-halo term of the projected mass profile
(Eq.~\ref{eq:lens_2h_app}). 
We note that we use the same parameters of the host halo,
($\bar{M}_{180b}, \bar{c}_{180b}$), when jointly fitting the model to the
measured mass profiles of the three centers, because the three
measurements used exactly the same catalog of multiple-LRG systems.
We do not assume a fixed value for the halo bias, instead
  treating $\bar{b}$ as a free parameter to absorb any
uncertainties in the modeling of this term (e.g., linear vs. nonlinear
power spectrum, linear bias only, etc.).  This is acceptable
  because even at the maximum scales shown on the plot, the 2-halo
  term is subdominant to the 1-halo term. 
We assumed a
narrow mass range of the host halos.  We also assume that all the ``Mean
centers'' in all the multiple-LRG systems
are off-centered (i.e. $q_{\rm cen}^{\rm Mean}=0$) and lack a 
sub-halo lensing contribution, because the Mean center is not associated
with the position of any bright galaxy.  
 Notice that we used the average
halo mass parameter, defined as the mass enclosed within a sphere of radius
$r_{180b}$ inside of which the mean density is 180 times the
mean background mass density $\bar{\rho}_{m0}$. 
We used the Markov Chain Monte Carlo (MCMC) method to explore
the marginalized posterior distribution of each model parameter, 
by taking into
account the significant correlations between the observed lensing profiles using
  different halo centers, and the
different radial bins. 

\begin{table*}
\begin{center}
Best-fit model parameters
\begin{tabular}{l|cccccccccc}
  \hline\hline
Measurement & $q_{\rm cen}^{\rm BLRG}$ 
& $R_{\rm off}^{\rm BLRG}$
& $m_{\rm sh}^{\rm BLRG}$
& $q_{\rm cen}^{\rm FLRG}$ 
& $R_{\rm off}^{\rm FLRG}$
& $m_{\rm sh}^{\rm FLRG}$
& $R_{\rm off}^{\rm Mean}$ 
& $\bar{M}_{\rm 180b}$
& $\bar{c}_{\rm 180b}$
\\
& $\!\!$[per cent]$\!\!$
& [Mpc/$h$]
& [$10^{12}M_\odot/h$]
& $\!\!$[per cent]$\!\!$
& [Mpc/$h$]
&  [$10^{12}M_\odot/h$]
& [Mpc/$h$]
& [$10^{14}M_\odot/h$]
&
\\
\hline
$\Delta\Sigma(R)$ & $63\pm 21$ 
& $0.44 \pm 0.22$ 
& $2.3\pm 1.4$
& $24\pm 13$
& $0.39 \pm 0.06$ 
& $1.9\pm 1.1$
& $0.16\pm 0.03$ 
& $1.63\pm 0.16$
& $4.8\pm 1.4$
\\
$w^{\rm cross}(R)$ & $54\pm 5$
& $0.35 \pm 0.02$ 
& --
& $32\pm 3$ 
& $0.40 \pm 0.01$ 
& --
& $0.19\pm 0.01$ 
& --
& -- \\
\hline
\end{tabular}
\end{center}
\caption{The best-fit values and marginalized errors of the model
 parameters
obtained from the fitting of the models to  measurement of the
 LRG-galaxy weak lensing  or, separately, the projected cross-correlation of LRGs and
photo-$z$ galaxies (Fig.\ref{fig:wl-wR_obs}). We focused on the
multiple-LRG systems for the measurements, and the error bar of each
parameter is the 1-$\sigma$ uncertainty including marginalization over
other parameters.
\label{tab:offset}
}
\end{table*}
Table~\ref{tab:offset} summarizes the parameter constraints derived from
the measured LRG-galaxy weak lensing. First of all, the host halo mass
for the multiple-LRG systems is about $1.6\times 10^{14}M_\odot/h$,
which is significantly more massive than the average halo mass inferred from
the LRG-shear lensing using all the LRG systems,
about $4\times10^{13}M_\odot/h$ \citep{2006MNRAS.372..758M} as
explicitly studied in Fig.\ref{fig:obs_single}.
Secondly, the central galaxy fraction
for the BLRGs is $q_{\rm cen}^{\rm BLRG}\simeq 63$ per cent, and larger than
$24$ per cent for the FLRGs. The measured mass profiles favor non-zero
$q_{\rm cen}$ values at the 
2 -- 3$\sigma$ level,
i.e. some contribution from central LRGs is needed to reproduce the measured
profiles.  We 
found non-zero values of the off-centering radius for all three
different centers (BLRGs, FLRGs and Mean): 
$R_{\rm off}\simeq 350,
400$ and 190 kpc/$h$ for the BLRG, FLRG and Mean centers, respectively,
although this detection is not very statistically 
significant for the BLRG center ($2\sigma$).  

The different panels of Fig.\ref{fig:dsigma} compare the best-fit model
predictions with the measured mass profiles for the three different
centers. It is clear that our model reproduces the measurements very well.
The mass profile for the Mean center has a clear signature
of dilution due to the off-centering effect.
This is  reasonable, because we do not
believe that the Mean position of LRGs
happens to coincide with the true center. For the lensing mass profiles for the
BLRG and FLRG centers, the best-fit model has two contributions
from the central LRGs and from the off-centered (satellite) LRGs.  The
sub-halo contribution of the BLRG or FLRG is a marginal detection, but
implies that the measured lensing profile needs such an additional
 mass concentration relative to the smooth NFW profile,
which can be caused by two effects: the 
sub-halo hosting the satellite LRGs,
or the 
baryonic concentration effect on the total mass around the central LRGs
\citep{Schulzetal:10}.  The sub-halo mass we measured 
could be due to these effects and therefore is difficult to
interpret physically; furthermore, the detection significance is quite
low, $<2\sigma$ for both BLRG and FLRG. 

Since we now have constraints on the
mass profile of host halos and the typical off-centering radius for
the different halo centers, we
can infer the typical velocity dispersion of the off-centered LRGs based
on the virial theorem \citep[][]{Hikageetal:12}:
\begin{equation}
\sigma^2_{{\rm off}, v}(r)=\frac{GM(<r)}{2r}.
\end{equation}
The enclosed mass within a sphere of radius $R_{\rm off}$ for an NFW
halo is given as $M(<R)=M_{180b}f(c_{180b}R/R_{180b})/f(c_{180b})$, where
$f(x)\equiv \ln(1+x)-x/(1+x)$ and $M_{180b}$, $R_{180b}$ and $c_{180b}$
are the virial mass, virial radius and the halo concentration for the
mass definition with respect to 180 times the mean mass density. As can
be found from Table~\ref{tab:offset}, if we assume $M_{180b}=1.6\times
10^{14}M_\odot/h$, $c_{180b}=4.8$, and the offset radii of 
$R_{\rm off}=440$ and $ 390~{\rm kpc}/h$
for the off-centered BLRGs and FLRGs,
 the virial theorem above gives 
$\sigma_{v,{\rm off}}\simeq 516 $ and $ 511~{\rm km}/$s, 
respectively. 
To be more precise, we estimate 
the typical velocity dispersion for the off-centered
LRGs  by weighting the velocity dispersion in the above
equation with the off-centering profile 
with width of $R_{\rm off}$ (Eq.~\ref{eq:poff}).
These velocity dispersions  are consistent with 
those inferred from the FoG measurements in
Fig.\ref{fig:fog} (which imply $\sigma_{v,{\rm off}}\simeq 500~$km/s).

Finally we comment on the significance of the detection of off-centered
BLRGs. As we discussed, the fraction of off-centered BLRGs, $q_{\rm
cen}^{\rm BLRG}=0.63\pm 0.21$, is less than a $3\sigma$ significance,
given the limited signal-to-noise ratios of the lensing profiles for the
multiple LRG systems (see Section~\ref{sec:conclusion} for a further
discussion). One might ask whether the measured lensing profiles for the
three centers can be fitted by requiring that all BLRGs sit at
the true center of each halo, i.e. $q_{\rm cen}^{\rm BLRG}=1$ and
$R_{\rm off}^{\rm BLRG}=0$. The best-fit $\chi^2$ is degraded only by
$\Delta \chi^2_{\rm min}=2$ compared to our fiducial model, and therefore
 a detection of the off-centered BLRGs is not significant,
less than a 2$\sigma$ level. 
In this case, the
best-fit values for other parameters are changed from 
Table~\ref{tab:offset}; 
$\bar{M}_{180b}
= 1.63 \rightarrow 
(1.59\pm 0.14)\times 10^{14}M_\odot/h $, 
$\bar{c}_{\rm 180b}=4.8 \rightarrow 3.3\pm 0.5$,
 $q_{\rm cen}^{\rm
FLRG}=0.24 \rightarrow 0.32\pm 0.15$, 
$R_{\rm off}^{\rm FLRG}=0.39 \rightarrow (0.37\pm 0.07)~{\rm Mpc}/h$, 
and $R_{\rm off}^{\rm Mean}=0.16\rightarrow (0.12\pm 0.02)~{\rm Mpc}/h$.  
It is worth noting that the halo concentration 
derived using 
the central BLRG assumption, $\bar{c}_{180b}=3.3$, is lower than that
predicted for halos of this mass by 
$N$-body simulations. 
When the lensing results are combined with the 
cross-correlations of the multiple-LRG halos with the photometric
galaxies, which have greater signal-to-noise ratios, we can derive more
robust, convincing constraints on the off-center parameters for the BLRGs
and FLRGs, as we will show below. 

\subsection{Projected cross-correlation between LRGs and photometric galaxies}

As the third observable to constrain the off-centered LRGs, we use the
projected cross-correlation function between the LRG-inferred halos 
and a purely photometric sample of galaxies, $w^{\rm cross}(R)$.  The sample used
for this calculation was selected in much the same way as the lensing
source catalog described in the previous section, but (1) there is no
requirement that the galaxy be large enough in angular extent to measure
its shape, (2) we impose a brighter flux limit of $r<21$ to reduce the
effects of photo-$z$ error, and (3) we
require that the ZEBRA fit to the galaxy SED be consistent with an
early-type galaxy and that the photo-$z$ is consistent with the LRG 
redshift to within the quoted $68$ per cent 
statistical errors from ZEBRA. 

Our measurement of the projected correlation 
between the
 LRG-inferred
 halos
($H$) and fainter photometric red galaxies ($G$) also makes use of a
sample of random points ($R$) distributed in the same way as the halos.
Given a lower and upper 68 per cent confidence limit on the faint
galaxy photo-$z$, $z_\mathrm{ll}$ and $z_\mathrm{ul}$, we compute the
  projected correlation via summation:
\begin{equation}
w^{\rm cross}(R) = \frac{\sum H G(z_\mathrm{ll}\le z_H \le z_\mathrm{ul})}{\sum R
  G(z_\mathrm{ll}\le z_R \le z_\mathrm{ul})} - 1,
\end{equation}
where the projected radius is estimated from the redshift of the LRG
halo and the angular separation between the halo center and the
photometric galaxy, $R=\chi_{\rm LRG} \Delta\theta$. 
The cross-correlation
method is very powerful to distinguish the photometric galaxies that are
physically associated and therefore clustered with the spectroscopic LRG
halos. Including photometric galaxies that are not associated with
LRGs, and appear at that redshift due to photo-$z$ error, causes
a dilution of the correlation signals. Hence, we will not 
use the amplitude information for the following parameter
estimation. Similarly to the LRG-galaxy lensing measurement, we
estimate the covariance matrix of $w^{\rm cross}(R)$ via the jackknife resampling
method.

The right panel of Fig.\ref{fig:wl-wR_obs} shows 
the measured cross-correlation, $w^{\rm cross}(R)$,
for the three halo centers. We again note that 
we worked on the same set of multiple-LRG
systems for the three centers, and the difference between the three
cross-correlation measurements
is only due to the positions of the LRG-inferred halo centers.
Interestingly, 
over the intermediate range of radii,
$R\simeq[0.15,0.5]~$Mpc/$h$,
the correlation functions have greater
amplitudes in the order of the Mean, BLRG and FLRG
centers. Then, for the larger radii, 
$R\ga 0.5$~Mpc/$h$,
the correlation
amplitudes are in the opposite order, greater in the order of the FLRG,
BLRG and Mean centers\footnote{While the magnitude of this effect
  appears small within the errors on the curves, the signals are
  highly correlated such that the statistical significance of these
  orderings is high as can be found from the lower panel of Fig.\ref{fig:wl-wR_obs}.}. 
As can be found from Fig.\ref{fig:off_model},
these behaviors are exactly what is expected for
the off-centering effects, 
because the off-centering effect causes a
transfer of the power of the correlation function from smaller
  to larger radii relative to the typical off-centering radius,
  {\em regardless} of the underlying true radial distribution of
  photometric galaxies in the LRG-inferred halos. 
The order of
the off-centering amounts for the BLRG, FLRG and Mean centers is in good
agreement with the FoG measurements in Fig.\ref{fig:fog}, and also
consistent 
 with the weak lensing measurements in the left panel of
Fig.\ref{fig:wl-wR_obs}.
The statement that the Mean center is,
 in a statistical sense, closer to the true center than the BLRG center
 for the multiple-LRG halos is favored by the relative
 amplitude differences between the BLRG and Mean centers in the 
different ranges of radii; 
 $w^{\rm BLRG}(R)>w^{\rm Mean}(R)$ in the range
$0.15\simlt R\simlt 0.5~{\rm Mpc}/h$,
 $w^{\rm BLRG}(R)<w^{\rm Mean}(R)$
in the range 
$0.5\simlt R\simlt 2~{\rm Mpc}/h$, and 
$w^{\rm BLRG}(R)\simeq w^{\rm Mean}(R)$ at the larger radii
as can be found from
Fig.\ref{fig:off_model}. 
On the other hand, the correlation function for the BLRG center
shows the greatest amplitude at the very small radii, $R\la
0.15~$Mpc/$h$, which is not as expected from the off-centering effect. We
will show below that the measurements can be nicely explained by
including 
 a mixture of central and satellite LRGs in the multiple-LRG
systems into the halo model predictions. 

\begin{figure*}
\begin{center}
\includegraphics[width=17.5cm]{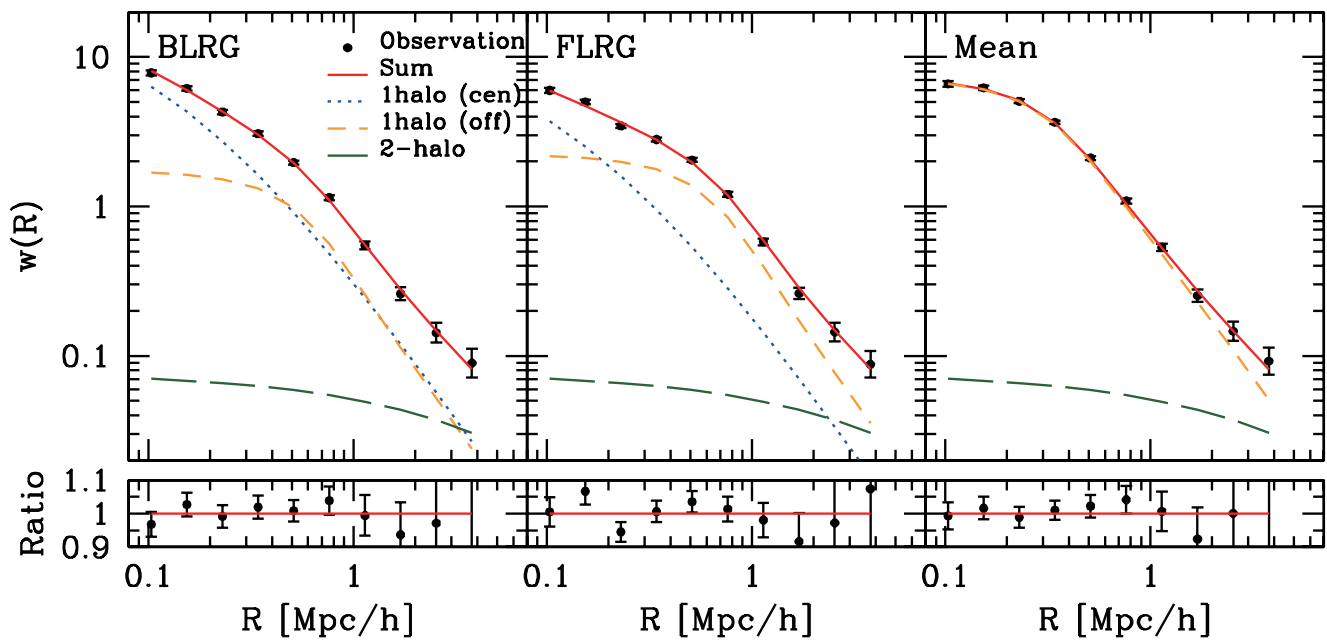} 
\caption{The best-fit model
predictions obtained by fitting the models (see
 Section~\ref{sec:angcor}) to  the
 projected cross-correlation functions of the LRG-inferred
halos with the photometric galaxies, for the BLRG ({\em left panel}),
FLRG ({\em middle}), and Mean ({\em right}) centers, respectively. As 
 in Fig.\ref{fig:dsigma}, we assumed that only some fraction of
 the BLRGs or FLRGs are off-centered galaxies.
By using a model containing a mixture of central and satellite
 LRGs, we can reproduce the measured projected correlation
 functions over the entire range of radii we consider. The lower plot
 in each panel shows a remarkably good 
 agreement between the best-fit model and the measured profile for each
 center. 
\label{fig:angcor}}
\end{center}
\end{figure*}

Similarly to the LRG-galaxy lensing, we fit the halo
model in Section~\ref{sec:angcor} to the projected correlation functions. 
To be more precise, we use the following model parameters:
\begin{equation}
p_{\alpha}=(\Delta z_{\rm phz}, \bar{M}_{180b}, \bar{c}^g_{180b},
q_{\rm cen}^{\rm BLRG}, R_{\rm
 off}^{\rm BLRG}, q_{\rm cen}^{\rm FLRG}, R_{\rm
 off}^{\rm FLRG}, 
R_{\rm off}^{\rm Mean},\bar{b}^2),
\label{eq:paras_wR}
\end{equation}
where $c^g_{180b}$ is the effective concentration parameter 
 to model
the radial distribution of photometric galaxies by an NFW model.
Most of the parameters are the same as those for the LRG-galaxy lensing
analysis (Eq.~\ref{eq:paras_wl}), except for the parameter $\Delta
z_{\rm phz}$ and $\bar{b}^2$. 
We again note that we employ the same parameters of the host halo, 
$(\bar M_{\rm
 180b}, \bar c_{180b})$, when jointing fitting the model to the
three cross-correlation functions for BLRG, FLRG and Mean centers,
because the three measurements used exactly the same catalog of 
 multiple-LRG systems and photometric galaxies.
The parameter $\Delta z_{\rm phz}$ is a nuisance parameter
to model the dilution factor, $\bar{n}^{\rm 2D}_{\rm phg}(z_{\rm
LRG})/\bar{n}^{\rm 2D}_{\rm phg, all}$, that 
arises from a contamination of unassociated photometric galaxies due to
photo-$z$ errors as we discussed above. 
We also include the bias parameter $\bar{b}^2$ to account for an
uncertainty in the 2-halo term amplitude (which is for
$\bar{b}_{\rm LRG}\bar{b}_{\rm phg}$ in Eq.~\ref{eq:wR_app}).
In our
fitting, we assumed a single (i.e. very narrow) mass bin for host
halos of the multiple-LRG systems, and that the distribution of photometric
galaxies in the halos follows an NFW profile (see below for the results
obtained by relaxing the NFW assumption).

Table~\ref{tab:offset} shows the best-fit value and the marginalized
$1\sigma$ errors for each parameter. The results are in remarkably good
agreement with the LRG-galaxy lensing results within the error bars,
while the two measurements are 
independent in a sense that they are subject to largely
  different systematic errors, and used different
catalogs of photometric galaxies: the background galaxies behind the
LRGs and the galaxies at the same redshift as the LRGs,
respectively. 

Fig.\ref{fig:angcor} shows the best-fit model predictions compared with
the measurements. The halo model can well reproduce the measured
correlation functions over all the range of radii we consider, including
the enhancement in the correlation amplitude at the small radii for the
BLRG center.  Even for the small error bars,
our best-fit model 
 gives a surprisingly good fit
to the three cross-correlation functions simultaneously, 
as explicitly shown in the lower panels for
each center. 

\begin{figure}
\begin{center}
\includegraphics[width=14cm]{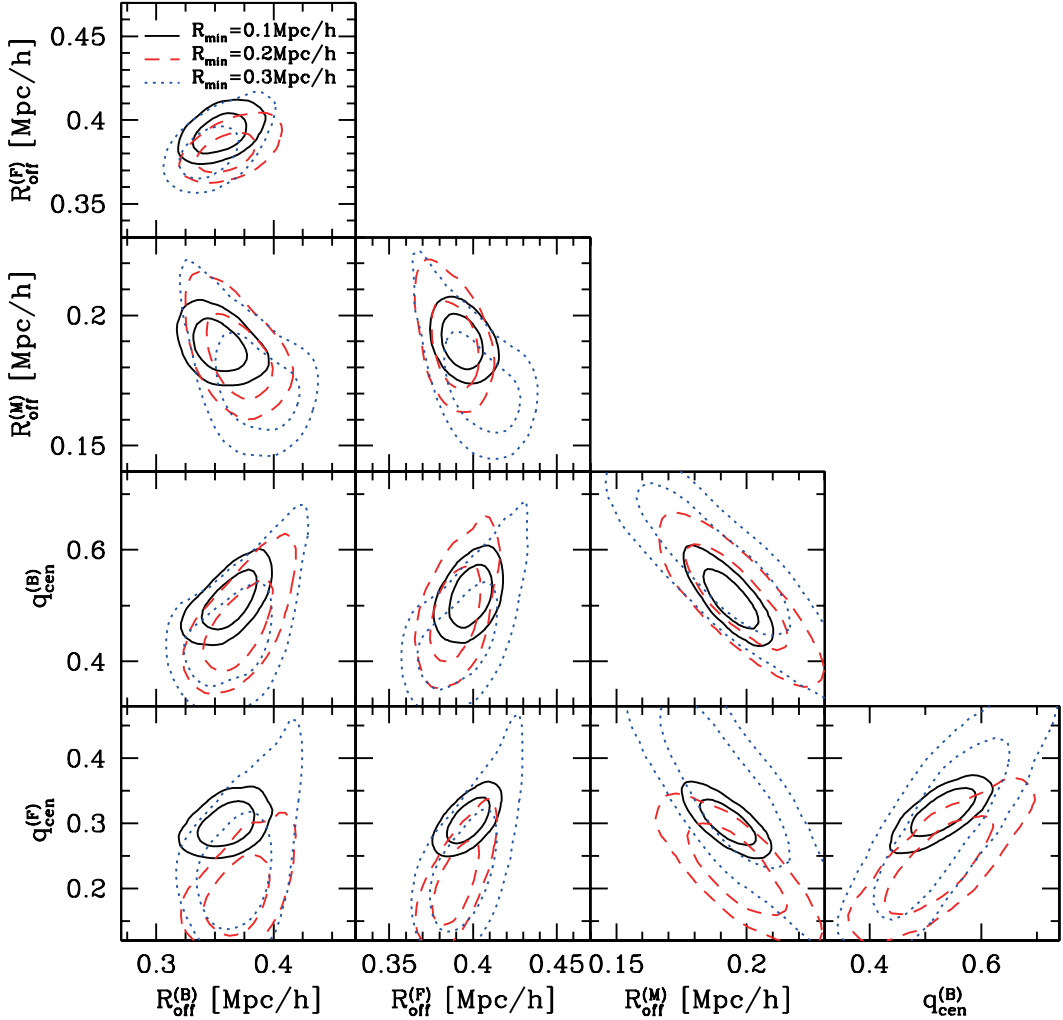}
\caption{The marginalized constraint contours (68 and 95 per cent C.L.) in
two-dimensional sub-spaces of the off-centering parameters. The dotted,
dashed and solid contours in each panel show the results when we
change the minimum radius for the fit  to the
cross-correlation functions of LRGs and photometric galaxies in
Fig.\ref{fig:angcor}, $R_{\rm min}=0.3, 0.2$ and $0.1~{\rm
Mpc}/h$. Including the smaller-radius cross-correlation
 helps to break the parameter degeneracies, especially the fraction of
 central LRGs, $q_{\rm cen}$. 
 \label{fig:2dpost_angcor} }
\end{center}
\end{figure}

\begin{figure}
\begin{center}
\includegraphics[width=14cm]{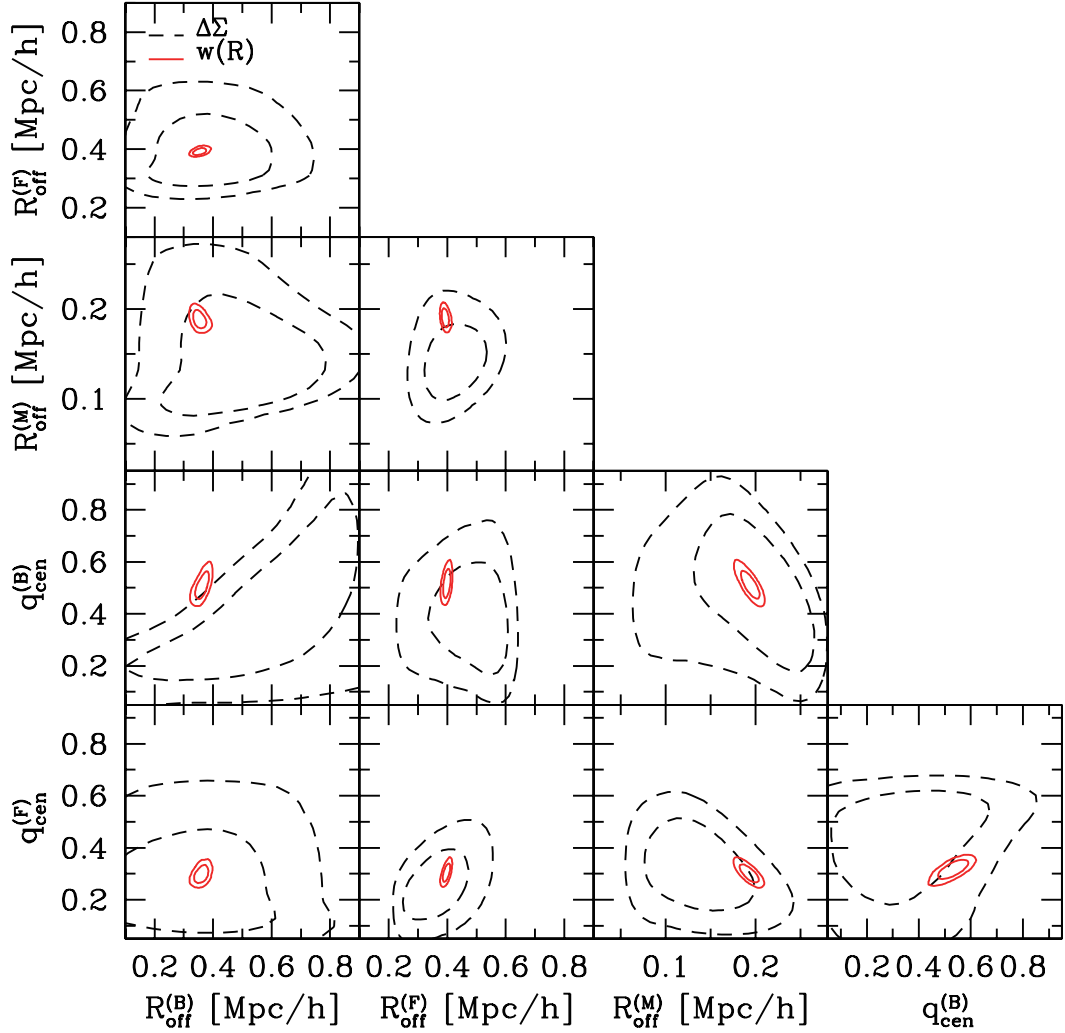}
\caption{Similar to the previous figure, but comparing the 68 and 95
  per cent C.L. marginalized constraint contours obtained
 from the LRG-galaxy weak lensing measurements (dashed contours) and the
 LRG-photometric galaxy cross-correlation (solid). 
\label{fig:2dpost_dsigma}
}
\end{center}
\end{figure}

In Fig.\ref{fig:2dpost_angcor}, we show the marginalized constraint
contours in each two-dimensional sub-space of the off-centering 
parameters. Comparing the dotted, dashed and solid contours shows how
the parameter degeneracies are broken by changing the smallest radius
down to which the cross-correlation signals are included in the model
fitting. It is clear that the enhancement in the cross-correlation
amplitudes at small radius, $R\simlt 0.3~{\rm Mpc}/h$, helps to break
parameter degeneracies, especially to constrain the central galaxy
fraction parameters $q_{\rm cen}$. 

Fig.\ref{fig:2dpost_dsigma} compares the constraint contours obtained
from the model fitting to the LRG-galaxy weak lensing and the
LRG-photometric galaxy cross-correlations. The figure shows that the two
results are in good agreement with each other within the errors, and
that the cross-correlation gives tighter constraints on the off-center
parameters than the weak lensing due to the higher signal-to-noise
ratios.  The difference in $S/N$ between the two types of
  measurements will likely be different in deeper upcoming lensing
  surveys than it is in SDSS.

One uncertainty in our halo model is the assumption that
the radial distribution of photometric galaxies 
follows an NFW profile. This does not necessarily hold for the SDSS
galaxies. In Appendix~\ref{sec:genNFW}, we study how the best-fit
off-center parameters are changed if we relax the NFW assumption. In
doing so, we used a generalized radial profile given by $n_g(r)\propto
1/[r^\alpha (1+r/r_s)^{\beta-\alpha}]$ and allow the additional
profile parameters $(\alpha, \beta)$ to freely vary in the model
fitting, in addition to $(\bar{M}_{
180b},\bar{c}^g_{180b})$ (hence, 2 additional free
parameters).  As shown in detail in Appendix~\ref{sec:genNFW}, we
found that the best-fit off-center parameters and their uncertainties
are almost unchanged.
To be more explicit, the best-fit
parameters are 
$q_{\rm cen}^{\rm BLRG}=0.53\pm 0.05, 
R_{\rm off}^{\rm BLRG}=(0.34\pm 0.02)~{\rm Mpc}/h,
q_{\rm cen}^{\rm FLRG}=0.32\pm 0.03$, 
$R_{\rm off}^{\rm FLRG}=(0.39\pm 0.01)~{\rm Mpc}/h,
$ and $R_{\rm off}^{\rm Mean}=(0.019\pm 0.01)~{\rm Mpc}/h$, which 
are compared with the values in Table~\ref{tab:offset}.
Thus we
conclude that the off-center parameters derived by our method are
relatively insensitive to the profile parameters, because the
constraints are derived by comparing the different cross-correlation
functions for different centers using the {\em same} halo and
photometric galaxy catalogs, as
well as by including marginalization over the profile parameters. 
We also stress that our results 
suggest that the radial distribution of photometric member
galaxies in cluster-scale halos of $\sim 10^{14}M_\odot/h$ is reasonably
well described by a generalized NFW profile with free concentration
parameter \citep[also
see][]{BerlindWeinberg:02,Moreetal:09,Chen:09,Guoetal:12}.  However,
we have not quantitatively explored the impact of more free models, so
the specific constraints that we quote for $q_{\rm cen}$ and $R_{\rm
  off}$ do contain some assumption regarding the profile of
photometric galaxies.  Since other studies have not suggested that
these distributions are, on average, pathological enough that they
cannot be described by a generalized NFW with free concentration, this
assumption is not in practice a significant weakness in our results.

\begin{figure}
\begin{center}
\includegraphics[width=8cm]{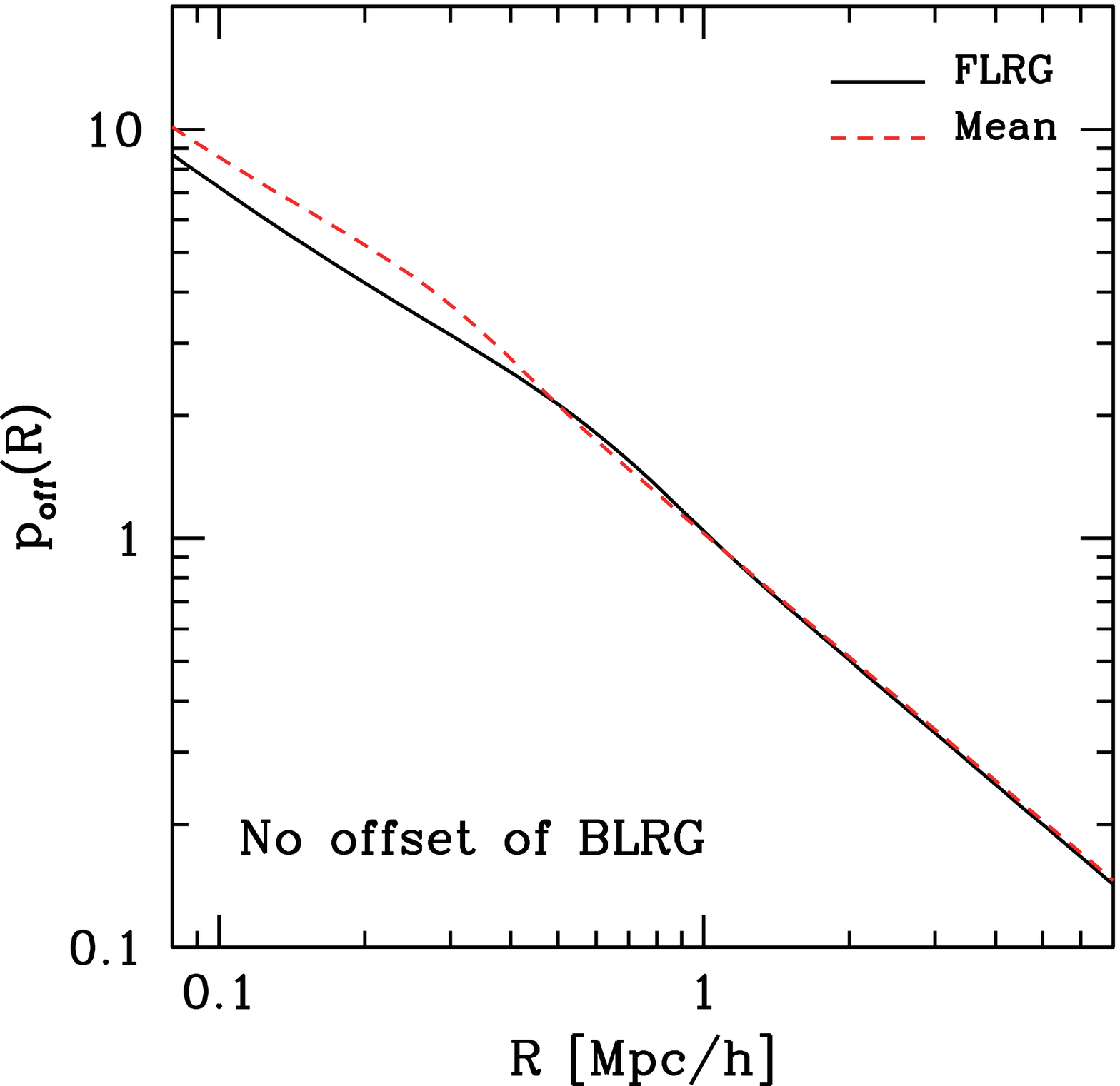}
\includegraphics[width=9cm]{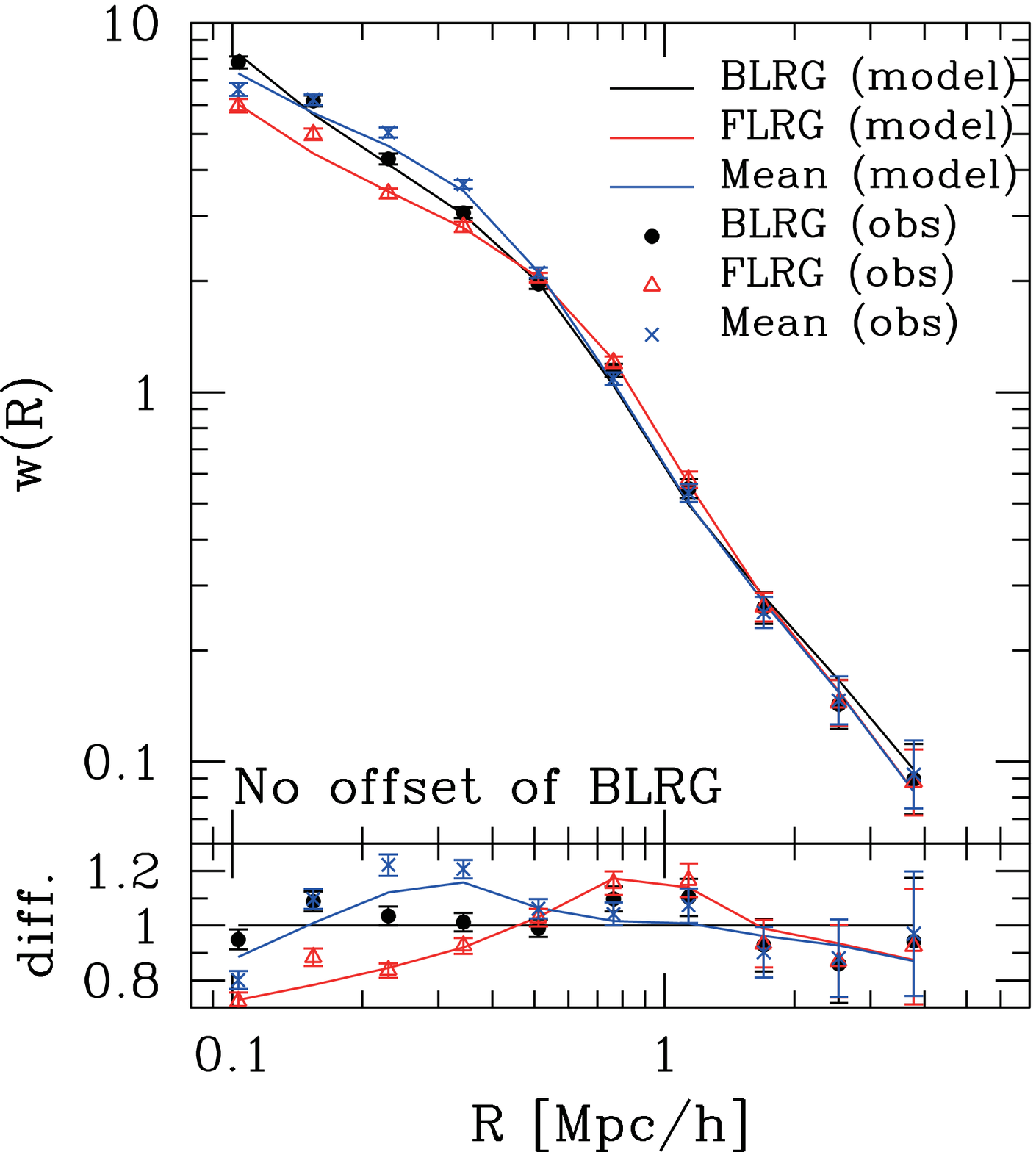}
\caption{A test of the central BLRG hypothesis for an
 interpretation of the cross-correlation measurements in
 Fig.~\ref{fig:angcor}, i.e. all the BLRGs are at the true center in each of
 the multiple-LRG system halos. Assuming that all the multiple-LRG
 systems are systems having 2 LRGs inside and therefore the Mean
 position in each host halo has a half of the off-center radius of
 FLRG, the radial (off-center) distribution of FLRG or Mean center
 proxy can be estimated by using the halo model with a generalized NFW
 profile to model the radial distribution of member galaxies (see
 text around Eq.~\ref{eq:cenBLRGhypothesis}
and Appendix~\ref{sec:genNFW} for details).
To include uncertainties in the
 model parameters allowed by 
the measurement errors, we obtained the best-fit
 model by 
varying the model
 parameters.
The $\chi^2$ value for the
 best-fit model, $\chi_{\rm min}^2=84$, can be compared to $\chi_{\rm
 min}^2=21$ for the best-fit model allowing $q_{\rm cen}^{\rm BLRG}\ne
 1$; therefore $\Delta \chi^2=63$, about
 8$\sigma$ deviation. {\em Left panel}: The normalized radial profiles of FLRG and
 Mean center proxies for the best-fit model with $q_{\rm cen}^{\rm BRLG}=1$. {\em Right panel}: The
 best-fit model profiles for BLRG, FLRG and Mean center proxies
 compared with the data (where the data points are the same as in the
 right panel of Fig.~\ref{fig:wl-wR_obs}). 
The lower panel shows the fractional difference of
 the best-fit model or the measured profile 
relative to the best-fit model profile for
 BLRG. The model profiles show sizable differences from the measured
 profile for each of the three centers over the different range of
 radii.    \label{fig:blrgnooff}}
\end{center}
\end{figure}
As for the lensing profiles in
  Section~\ref{sec:weaklens}, one might wonder whether the measured
  cross-correlation functions can be interpreted by imposing all the
  BLRGs to be at the true center in each halo, $q_{\rm cen}^{\rm BLRG}=1$ and
  $R_{\rm off}^{\rm BLRG}=0$.  
If we employ 
this hypothesis, we can realize that the BLRG
cross-correlation, $w^{\rm BLRG}(R)$, gives the average radial
distribution of photometric galaxies in the multiple-LRG halos, while
the FLRG profile, $w^{\rm FLRG}(R)$, arises from a convolution of the
radial distribution of photometric galaxies with the radial distribution
of FLRGs in the same host halos.  Since our halo model assuming a
generalized NFW profile ($n_g(r)\propto 1/[r^\alpha(1+r/r_s)^{\beta-\alpha}]$)
gives a good fit to the
measured profiles, we can use the
1-halo term profiles to infer the radial profiles of the photometric
galaxies and FLRGs:
\begin{eqnarray}
&&\tilde{u}_g(k)
=q_{\rm cen}^{\rm BLRG}\tilde{u}_{\rm gNFW}(k)
+(1-q_{\rm cen}^{\rm BLRG})\tilde{p}_{\rm off}^{\rm BLRG}(k; R_{\rm off}^{\rm BLRG})
\tilde{u}_{\rm gNFW}(k),
\nonumber\\
&&\tilde{u}_g(k)\tilde{P}_{\rm off}^{{\rm FLRG}}(k)
=q_{\rm cen}^{\rm FLRG}\tilde{u}_{\rm gNFW}(k)
+(1-q_{\rm cen}^{\rm FLRG})\tilde{p}_{\rm off}^{\rm FLRG}(k; R_{\rm
off}^{\rm FLRG}).
\label{eq:cenBLRGhypothesis}
\end{eqnarray}
The terms on the r.h.s. of the above equations are our models allowing
the off-centered BLRGs: $q_{\rm cen}^{\rm BLRG}$ or $q_{\rm cen}^{\rm
FLRG}$ is the fraction of central BLRGs or FLRGs, $\tilde{p}_{off}^{\rm
BLRG}$
or $\tilde{p}_{\rm off}^{\rm FLRG}$ is the Fourier transform of the
radial distribution of off-centered FLRGs or BLRGs,
 and $\tilde{u}_{\rm
gNFW}(k)$ is the Fourier transform of the generalized NFW profile to
model the radial distribution of member galaxies, which is given by 4
free parameters ($M_{180b}, c_{\rm 180b}^g, \alpha, \beta$) (see
Appendix~\ref{sec:genNFW} for details).  The terms on the l.h.s are from
the central BLRG hypothesis: $\tilde{u}_g(k)$ is the Fourier transform
of the radial distribution of photometric galaxies, and $\tilde{P}_{\rm
off}^{\rm FLRG}(k)$ is the Fourier transform of the radial distribution
of FRLGs. In this way, we can derive $\tilde{u}_g(k)$ and
$\tilde{P}_{\rm off}^{\rm FLRG}(k)$ in Fourier space. Then, assuming
that all the multiple-LRG systems are systems with 2 LRGs inside, which
is a good approximation as found from Table~\ref{tab:lrgs}, the radial
profile of the Mean centers can be simply obtained as $\tilde{P}^{\rm
Mean}_{\rm off}(k)\simeq \tilde{P}_{\rm off}^{\rm FLRG}(2k)$, where
``2'' in the argument arises because we stretch the FLRG profile by a
factor of 2 in Fourier space to obtain the Mean profile (shrink the
radial profile by a factor of 2 in real space). Thus, by inverse
Fourier-transforming the product $\tilde{u}_g(k)\tilde{P}_{\rm off}^{\rm
Mean}(k)$, we can obtain the radial profile for Mean centers predicted
from the central BLRG hypothesis.
To include the effect of possible variations in the model parameters
within the measurement errors, we searched the best-fit model by varying
the model parameters (the parameters on the r.h.s. of
Eq.~\ref{eq:cenBLRGhypothesis}).
Fig.~\ref{fig:blrgnooff} shows the best-fit model predictions for the
radial profiles of FLRG and Mean center proxies and the
cross-correlation profiles for BLRG, FLRG and Mean center proxies,
respectively.  The minimum $\chi^2$ value for the best-fit model,
$\chi_{\rm min}^2=84$, should be compared to $\chi^2_{\rm min}=21$ for the
best-fit model that allows $q_{\rm cen}^{\rm BLRG}\ne 1$.
 Compared to the lower panels in
Fig.~\ref{fig:angcor}, the model profiles show sizable differences from
the measured profiles for each of the BLRG, FLRG and Mean center
proxies. Although the reconstruction method relies on the assumption
that the radial profiles for the photometric galaxies and FLRGs are
described by our adopted functional forms (generalized NFW for the
former and Gaussian profile for the latter) even outside the 
range of radii in the measurements, the results again disfavor the central
BLRG hypothesis. 

\section{Implication for the residual FoG effect in the LRG
 power spectrum}
\label{sec:fog_residual}

\begin{figure}
\begin{center}
\includegraphics[width=6.cm]{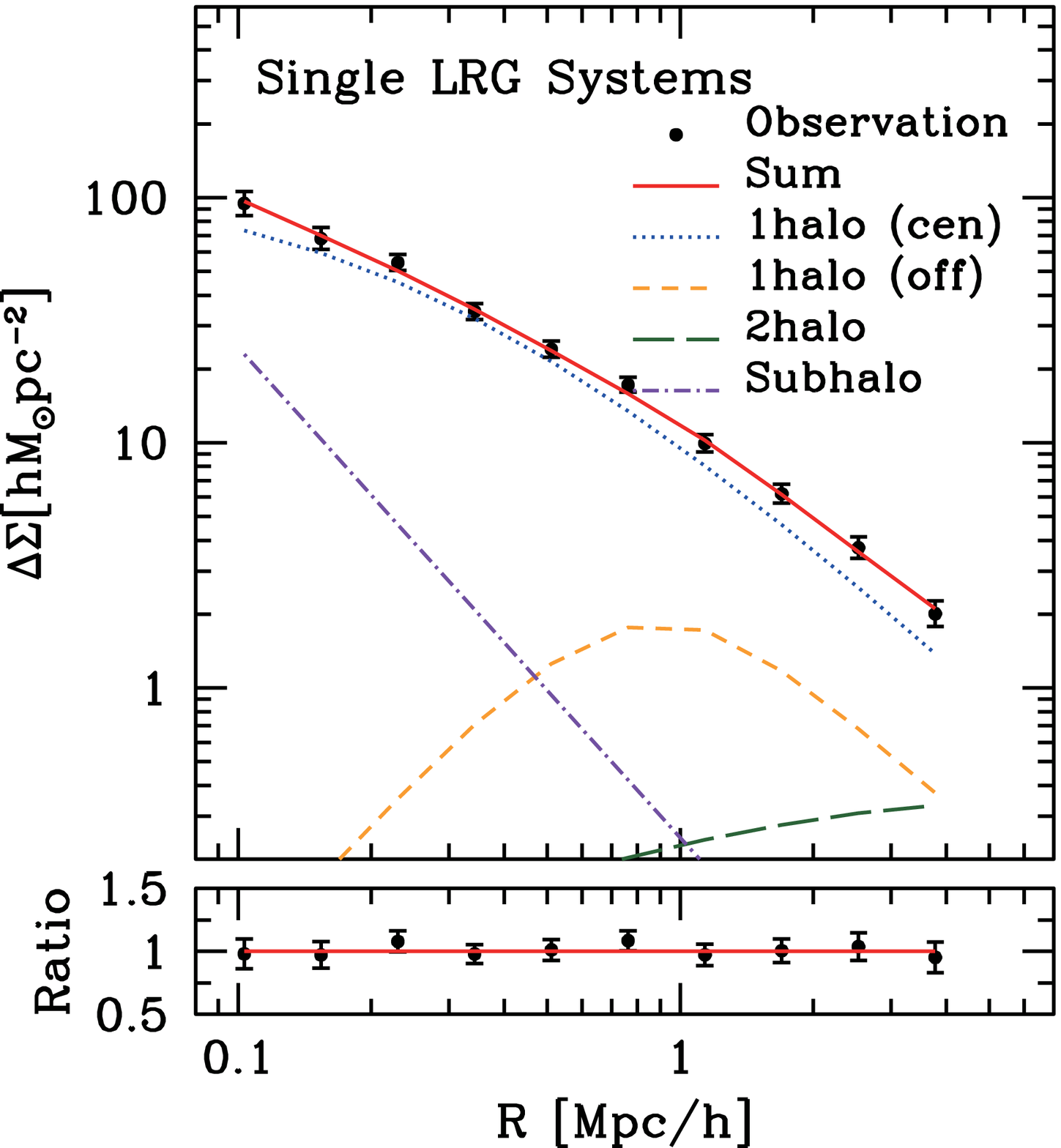}
\includegraphics[width=6.cm]{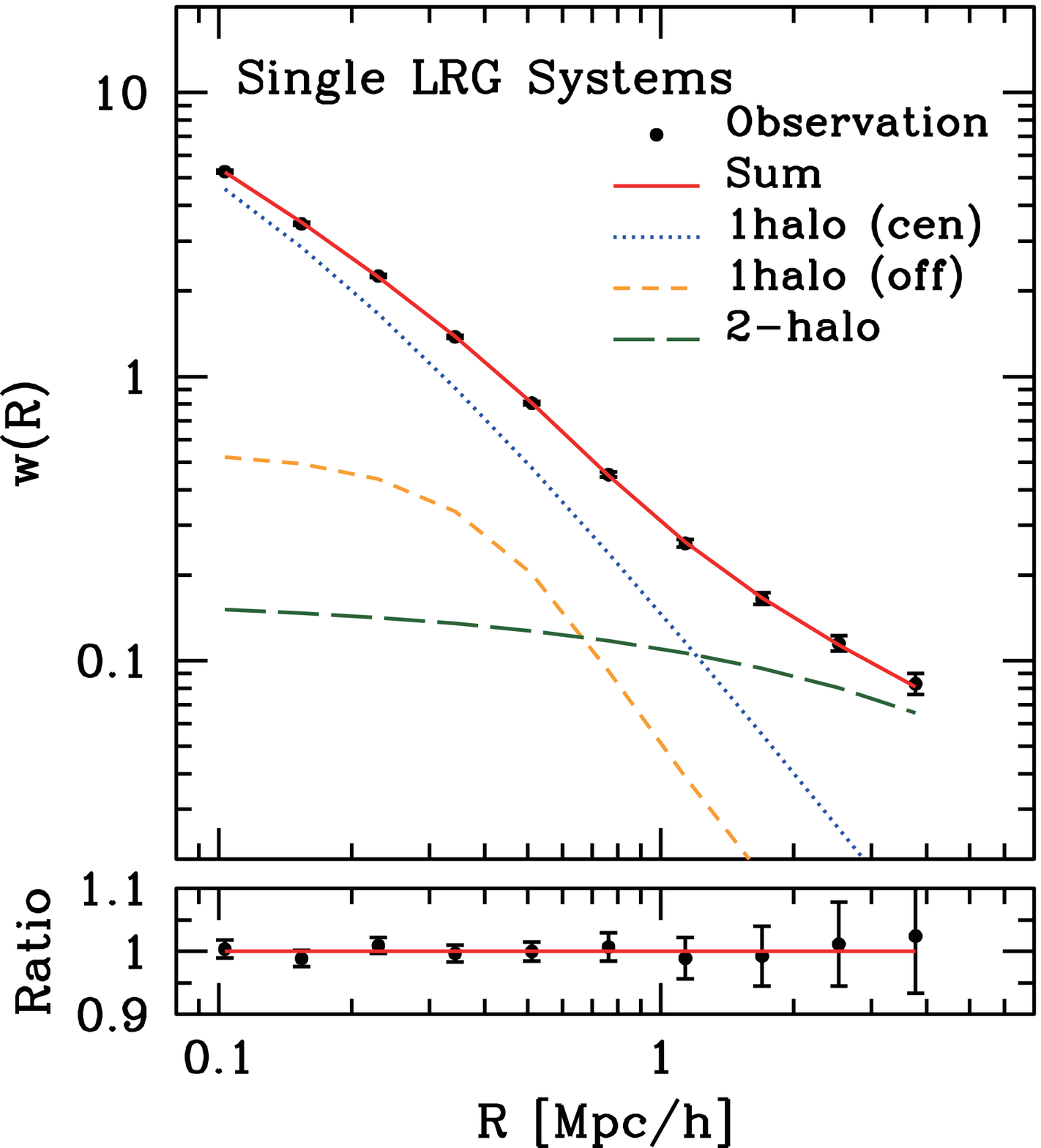}
\includegraphics[width=5.4cm]{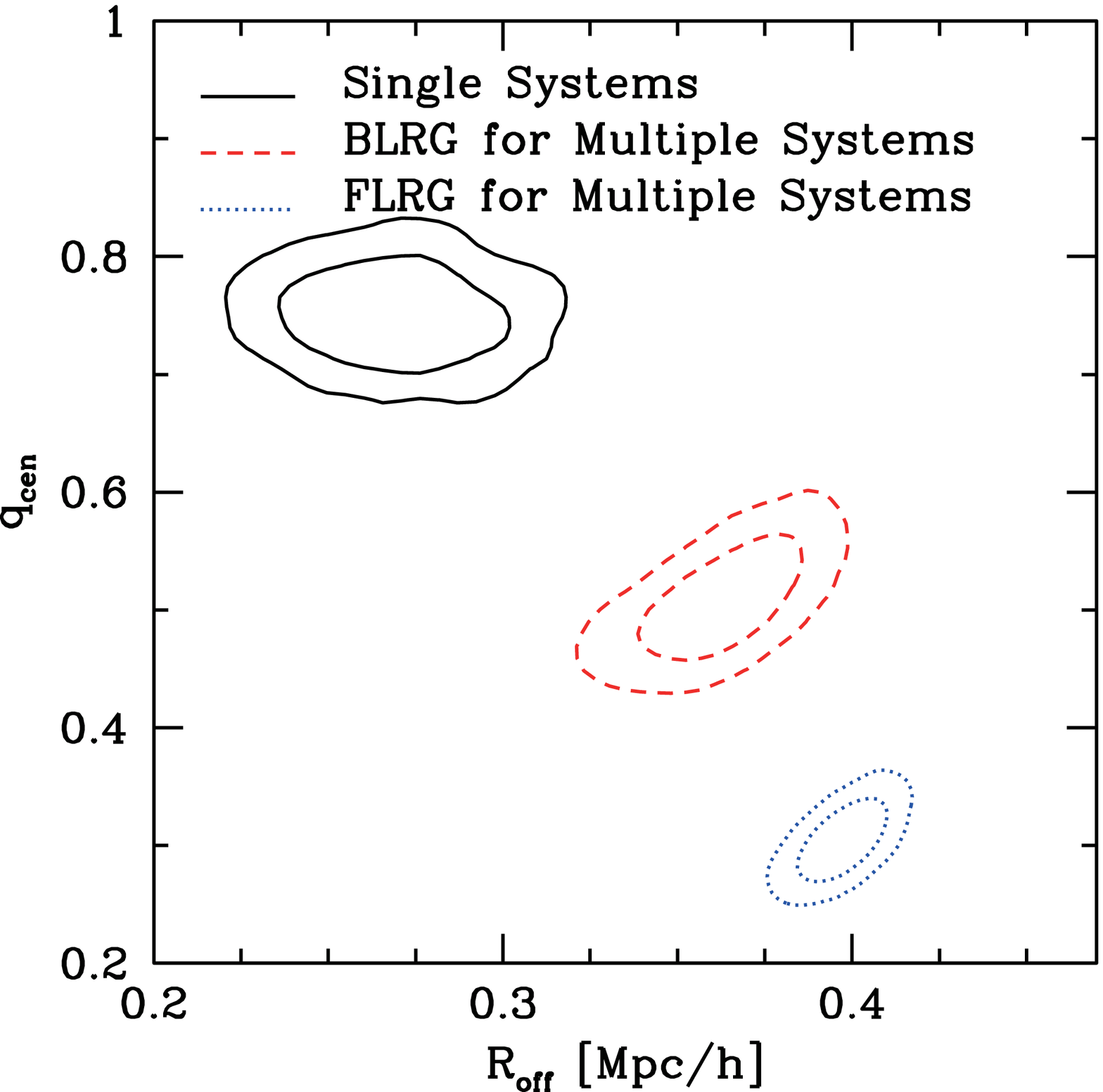}
\caption{The measured LRG-galaxy weak lensing ({\em Left panel}) and the
cross-correlation of LRGs with photometric galaxies ({\em Middle}) for
the single LRG systems. In each panel, the different curves show the
best-fit halo model predictions including the contributions from the
centered and off-centered LRGs, as in Figs.~\ref{fig:dsigma} and
\ref{fig:angcor}. However, in the model fitting we employ several 
simplified assumptions for the fitting
for simplicity: we assume a narrow mass bin for the host halo and a
thin redshift bin for the single LRG systems, 
and assume that the dark matter and the photometric galaxies in the
host halos follow an NFW profile. The LRG-galaxy lensing gives
$M_{180b}\simeq (0.42\pm 0.04)\times 10^{14}M_\odot/h$ for the best-fit
average mass for the host halos, which is smaller than that for the multiple-LRG
systems, $M_{180b}\simeq 1.6\times 10^{14}M_\odot/h$ 
(see Table~\ref{tab:offset}). {\em Right panel}:  the
best-fit off-centering parameters obtained from the cross-correlation
measurement are $q_{\rm cen}=0.76\pm 0.06$ 
and $R_{\rm off}=(0.26\pm 0.03)~{\rm Mpc}/h$ for the fraction of
central LRGs and the typical off-center radius, respectively.  A non-zero
off-centering effect is detected at $>2\sigma$ significance,
and the amount is smaller than in the multiple-LRG systems.  Note that
the best-fit parameters for the LRG-galaxy lensing are consistent with
the cross-correlation results within the error bars, but have larger
error bars.  \label{fig:obs_single}}
\end{center}
\end{figure}

We have so far shown that three different measurements (the
angle-averaged redshift-space power spectrum, the LRG-shear correlation
function and the cross-correlation of LRGs and photometric galaxies), 
all imply
the existence of satellite (off-centered) LRGs for the multiple-LRG
systems, which represent $\sim 5$ per cent of the halos hosting
LRGs. Perhaps most surprising is that some of BLRGs in the multiple-LRG
systems exhibit this off-centering effect.
In this section, based on these results,
we discuss the implications for possible residual FoG
contamination of the LRG power spectrum.  
For some of the multiple-LRG systems, the BLRG is
brighter than the FLRG only by
a few tenths of a magnitude (see Fig.\ref{fig:LRGs_dm}). One might
think that the off-centered BLRGs are mainly from the multiple-LRG
systems where the BLRG and FLRG have a small magnitude difference and
therefore the two LRGs are not different (the FLRGs are central
instead). In Appendix~\ref{sec:lrg_dm}, by dividing the multiple-LRG
systems into halved samples which have the larger and smaller magnitude
difference between the BLRG and FLRG than the median, we study whether
or not the amount of the off-centered BLRGs differs in between the
halved samples.  We do not find any significant difference for the
off-centered BLRGs in the two samples, although the FLRG center shows a
significant difference. Therefore, the current data show some 
off-centering effects for BLRGs in the multiple-LRG systems, 
irrespective of the magnitude difference between the BLRG and FLRG.

The single-LRG systems, each of which contain only one LRG (see
Table~\ref{tab:lrgs}), constitute about 95 percent of the LRG-inferred
halos. Although the LRG in single-LRG systems might have the
off-centering effect, we do not have any other proxy of halo center
besides the LRG position, unlike in the multiple-LRG
systems. Nevertheless, we can estimate the amount of possible
off-centering effect as follows.  

Fig.\ref{fig:obs_single} shows the measured LRG-galaxy lensing
and the measured cross-correlation of LRGs with photometric galaxies for
the single-LRG systems. 
As in Figs.~\ref{fig:dsigma} and \ref{fig:angcor}, we compare the halo
model with the
measured correlations, and show the 
 best-fit halo model predictions
including
contribution of the off-centered LRGs.
In this model
fitting, we employ several simplified assumptions;
we assumed a narrow mass bin for the host halos as well as a thin
redshift bin of LRGs, and assumed that the dark matter and the
photometric galaxies in the single-LRG systems both follow an NFW
profile. 
The LRG-galaxy lensing gives 
$M_{180b}\simeq
(0.42\pm 0.04)\times 10^{14}M_\odot/h$ for the best-fit host halo mass,
which is less massive than the host halo
mass of $M_{180b}\simeq 1.6\times 10^{14}M_\odot/h$ for the multiple-LRG
systems.  
The cross-correlation gives non-zero off-centering parameters: 
 $q_{\rm cen}=0.76\pm 0.06$ and 
$R_{\rm off}=(0.26\pm 0.03)~{\rm Mpc}/h$. The LRG-galaxy
lensing gives consistent results, but with larger error bars due to the
smaller signal-to-noise ratio.
However, the constraints on the off-centering parameters are not
as robust as those for the multiple-LRG systems, because we cannot
compare the measurements for the different halo center proxies and the
offsetting parameters are more degenerate with the NFW profile
parameters. If we use a generalized profile, $\rho(r)\propto
1/[r^\alpha(1+r/r_s)^{\beta-\alpha}]$, and allow the two additional
parameters ($\alpha,\beta$) to vary, the constraints on the off-center
parameters are relaxed to $q_{\rm cen}=0.79\pm 0.08$ and $R_{\rm
off}=0.26\pm 0.15~{\rm Mpc}/h$. The non-zero off-center radius 
becomes less significant, less than 2$\sigma$ level. Nevertheless, as
shown in the lower panel, the best-fit model can well reproduce the
measured correlation functions, within the small error bars over the 
range of radii we consider.  The LRG-galaxy lensing profile does not
give as tight constraints on the off-center parameters as 
the cross-correlation, due to the smaller signal-to-noise
ratios. Assuming an NFW profile, we obtained $q_{\rm cen}=0.76\pm
0.18$ and $R_{\rm off}=0.38 \pm 0.25~{\rm Mpc}/h$,
  which is consistent with the constraints from the cross-correlation
  (also see Appendix~\ref{sec:fog_wl}). We again note that the results
  in Fig.\ref{fig:obs_single} suggest that the radial distribution
  of member galaxies in group-scale halos of $\sim 4\times
  10^{13}M_\odot/h$ is fairly well fitted by a generalized NFW profile
  with free concentration.

\begin{figure}
\begin{center}
\includegraphics[width=10.6cm]{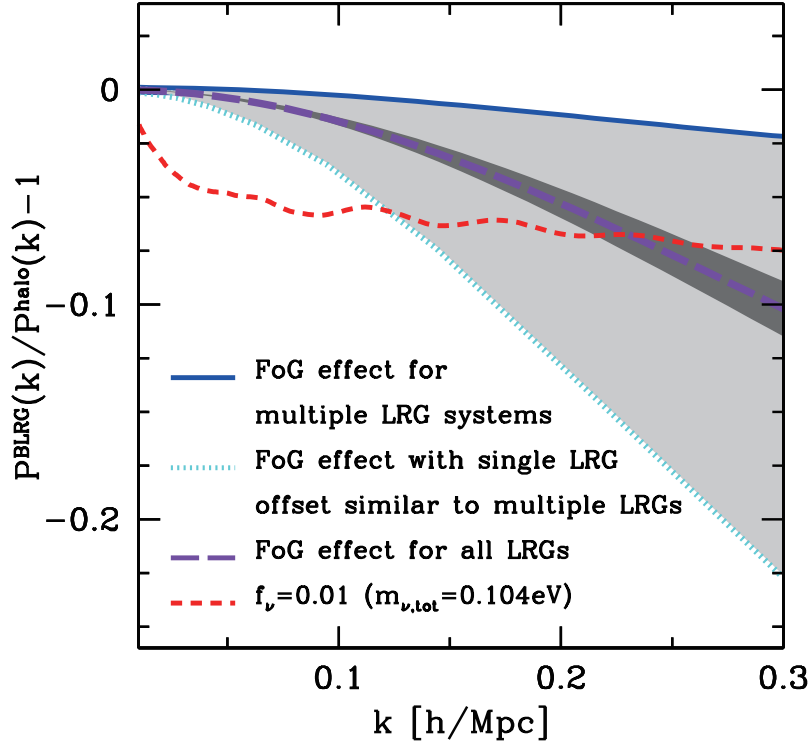}
\caption{Expected FoG suppression of the angle-averaged, redshift-space
power spectrum of LRGs  using the BLRG center
proxy. Here, we show a possible range of the residual FoG contamination
relative to the underlying halo power spectrum. 
The upper solid curve of the shade region
shows the FoG effect assuming that only BLRGs in the multiple-LRG
systems have the off-centering effects, as in Fig.\ref{fig:fog}.  The lower
dotted curve shows the FoG effect assuming that BLRGs in the single LRG
systems also have off-centering effects, computed based on the halo model
in \citet{Hikageetal:12}. For this case, we assumed that the off-centering
 radius scales with the virial radius of host halos 
as $R_{\rm off}\propto r_{180b}(M)$, and the FoG amplitude 
is  normalized so
 as to reproduce the FoG effect 
at the host halo mass of the multiple
 LRG systems, $M_{180b}=1.67\times 10^{14}M_\odot/h$. For the fraction
 of central BLRGs in each halo, we assumed that it varies with host halo
 mass as $q_{\rm cen}^{\rm
BLRG}=0.6+0.05\ln(M/M_{180b, {\rm WL}})$, and normalized to $q_{\rm
cen}=0.6$ at $M_{180b}=1.6\times 10^{14}M_\odot/h$. Thus the dotted
curve gives the worst case scenario for the residual FoG effect (see
 text for details).
The bold dashed curve is the FoG effect implied from the constraints on
 off-centering parameters for the single-LRG systems shown in
 Fig.\ref{fig:obs_single}, and the dark shaded region around the curve
 implies the range covered by $1\sigma$ uncertainties of the 
off-centering parameters. 
For comparison, the dashed curve shows the suppression effect
 caused by massive neutrinos with total mass 
  $f_{\nu}=0.01$ ($m_{\nu, {\rm tot}}=$0.125~eV). This reveals that
  the 
 possible FoG effect suggested by the results in this paper can be a
 serious source of systematic errors in estimating
 cosmological parameters using the LRG power spectrum without
   directly modeling the effect. 
\label{fig:pk_fog}
}
\end{center}
\end{figure}
In
Fig.\ref{fig:pk_fog}, we estimate the  
 possible residual FoG contamination in 
the LRG power spectrum. Shown here is the angle-averaged redshift-space
power spectrum for the BLRG center relative to the underlying true halo power
spectrum, where we have assumed the linear
Kaiser redshift-space distortion effect. Here the shade region shows 
the allowed range of the residual
FoG effect according to the measurements shown in this paper. 
The upper solid curve is the
FoG effect if only BLRGs in the multiple-LRG systems (and
  not any of the single-LRG systems) have the
off-centering effects, i.e. the FoG effect, as we found in
Fig.\ref{fig:fog}\footnote{One might notice that 
the amplitude of the FoG suppression in
Fig.\ref{fig:pk_fog} differs from 
 that in Fig.\ref{fig:fog}. This is  because 
we show the fractional difference of the BLRG power spectrum 
relative to the spectra for the Mean center
or the halo center in  
Figs.~\ref{fig:fog} and \ref{fig:pk_fog}, respectively. 
Since the power spectrum
for the Mean-center also has the FoG suppression as we have shown,  
the FoG effect for the
BLRG spectrum appears to be larger, when plotted with respect to 
 the spectrum for the halo center in Fig.\ref{fig:pk_fog} than to the
 spectrum for the Mean center in Figure\ref{fig:fog}.}. 
Hence, this is a lower limit on the residual FoG
effect that should exist in the BLRG power spectrum. 
The lower dotted curve is the worst-case scenario, which shows the FoG
contamination if BLRGs in the single-LRG systems also have the
off-centering effects in the host halos. 
To compute this
curve, we used the method developed in
\cite{Hikageetal:12}. To be more precise, we assumed that the off-centering
radius $R_{\rm off}(M)$ scales with the virial radius of host halos
 as $R_{\rm
off}(M)\propto r_{180b}(M)$, and determined the proportional coefficient
so that it reproduces the measurement results for the host
halo mass of multiple-LRG systems, $M_{180b}\simeq 1.6\times
10^{14}M_\odot/h$. In addition, we assumed that the fraction of central
BLRGs  scales with the halos mass as $q_{\rm cen}^{\rm
BLRG}=0.6+0.05\ln(M/M_{180b, {\rm WL}})$, where the normalization is
determined so as to give $q_{\rm cen}^{\rm BLRG}=0.6$ at $M_{180b,{\rm
WL}}=1.6\times 10^{14}M_\odot/h$,  and the slope $0.05$
is taken from the mock galaxy catalog used in
\cite{Johnstonetal:07}. 
The genuine FoG effect in the
BLRG power spectrum would be inside
the shaded region. 
The thick-dashed curve shows the FoG effect implied from the correlation
measurements of the single-LRG systems  in
Fig.\ref{fig:obs_single}. The dark shaded region around that curve
denotes the range covered by varying the off-centering parameters within
the $1\sigma$ errors. This result might be closer to the
genuine effect. However, since
the off-centering constraints for the single-LRG systems are derived by
employing several additional assumptions as we mentioned above,
we believe that the entire range within the shaded gray region is still allowed.

Thus, Fig.\ref{fig:pk_fog} implies that the possible FoG effect can be
 a systematic error in estimating
cosmological parameters if the residual FoG effect is ignored in the model
 fitting. 
For example, we show 
the effect of massive neutrinos on the halo power
spectrum  as the dashed curve (assuming
that the sum of 
neutrino masses is $0.125$~eV, close to the lower bound of the inverted
mass hierarchy), but fixing other cosmological parameters to the fiducial
values. Since the massive neutrinos  cause a suppression in the measured
power spectrum, as does the FoG effect, 
 the figure implies that neglecting the FoG effect
might cause a serious systematic error in the derived neutrino
mass, with the severity of the issue depending on what scales
  are used.

\section{Discussion and Conclusions}
\label{sec:conclusion}

In this paper, we combined three observables (the redshift-space LRG
power spectrum, LRG-galaxy weak lensing, and the projected
cross-correlation of LRGs and photometric galaxies) to 
constrain the fraction of satellite LRGs and the average off-centered
distribution of the satellite LRGs in the host halos.  When doing so, 
we focused on multiple-LRG systems, about 4000 systems in total
(4.5 per cent of all LRG-inferred halos, see
Table~\ref{tab:lrgs}), and compared the measurements obtained by using
the different halo center proxies: the positions of the brightest LRG (BLRG),
the faintest LRG (FLRG), and the mean positions (Mean) in each multiple-LRG system (Figs.~\ref{fig:fog} and \ref{fig:wl-wR_obs}). All three
observables consistently imply that some fraction of LRGs
are satellites -- even those that are the brightest in the halo -- and
have offsets from the true center. Among the three 
measurements, the projected cross-correlations of the LRG-inferred halo
centers with photometric galaxies show the strongest evidence for 
off-centered LRGs, even when allowing the satellite galaxies to
  follow a generalized NFW profile with four free parameters.  Perhaps most intriguing or surprising is that some
of the 
BLRGs are satellites and have offsets from the halo center, and the
Mean positions tend to be
closer to the true center, showing the smallest off-centering effect.

By comparing the halo model developed in Section~\ref{sec:model} with
the measurements, we constrained the model parameters including the
fraction of satellite LRGs and the average off-centering radius
(Table~\ref{tab:offset}, and Figs.~\ref{fig:fog}, \ref{fig:dsigma} and
\ref{fig:angcor}). A brief summary of our findings is as follows. About
60 and 30 per cent of the BLRGs and FLRGs, respectively, are central galaxies
in the host halos of the multiple-LRG systems, and the remaining LRGs
are satellite LRGs. If we do not employ a mixture of central and
satellite LRGs, we cannot self-consistently model the measurements
using different choices for the halo 
centers. The typical off-centering radius is about 400~kpc/$h$ for both the
satellite BLRGs and FLRGs. We also confirmed that the Mean centers just
{\em happen to } be statistically closer to the true center than the
BLRG- and FLRG-preferred centers. The LRG-galaxy weak lensing for the
Mean centers shows a clear dilution effect on small scales.
A similar off-centering
radius for the satellite BLRGs and FLRGs might be reasonable, because
the FLRGs and BLRGs are both luminous, early-type galaxies, and
therefore should have a similar age and history of
dynamical friction in the host halos; moreover, the luminosity gap is
typically small. The only difference in the
off-centering effects arises from the different fractions of central
LRGs among the BLRG and FLRGs. Even if using a
generalized NFW profile with free concentration, inner and outer slope
parameters to model the radial distribution of photometric galaxies, 
the significance of our detection of satellite BLRGs is not largely degraded.
Our results also suggest that the radial distribution of photometric
galaxies in cluster- and group-scale halos is fairly well fitted by a
generalized NFW profile with free concentration, though the exact
values we find (and their errors) are dependent on this assumed
distribution for the radial distribution of photometric galaxies.

In these constraints, weak lensing plays a
very complementary and unique role, although the
signal-to-noise ratio is low due to the small number density of
background galaxies in the shallow SDSS catalog. The mean mass of the
host halos of the multiple-LRG systems is constrained to be about
$1.6\times 10^{14}~M_{\rm 180b}/h$ (with a roughly 10 per cent
statistical error); this is indeed more massive than
the mean halo mass of all the LRGs. 
Such large offsets of bright galaxies in massive halos was also
suggested recently by \cite{Sehgaletal:12}, based on a study of the
stacked 
Sunyaev-Zel'dovich (SZ) effect via a cross-correlation of the SDSS
optically-inferred clusters with the CMB map of the Atacama Cosmology
Telescope survey.  They noted that one explanation for the unexpectedly smaller SZ
signal is large off-centering effects. It would be worth
exploring whether our results can resolve the stacked SZ result.

Based on these constraints, we explored the off-centering effects on the underlying redshift-space power
spectrum of LRG-inferred halos, and showed the allowed range for the residual FoG
contamination in the LRG power spectrum (Fig.\ref{fig:pk_fog}). As the most
optimal case, the residual FoG effect is only from the multiple-LRG
systems, as we have studied in this paper. As the worst case scenario,
by assuming that  a majority of LRGs in the (far more numerous) single-LRG systems
have similar off-centering effects as in the multiple-LRG
systems,  we estimated the possible maximum, residual FoG
effect on the LRG-inferred power spectrum, based on the method in our
earlier paper \citep{Hikageetal:12}. The genuine amount of residual
FoG effect should be between the two cases we show. In particular, the
residual FoG effect, if not accounted for, may cause a
potentially significant bias in
cosmological parameters derived from the measured power spectrum
amplitude, e.g. the sum of neutrino masses which likewise causes a suppression in
the power spectrum amplitude at small scales. Our results in Fig.\ref{fig:pk_fog} imply that the
bias may not be that significant if using the power spectrum amplitude only up to
$k_{\rm max}\simeq 0.1~h/$Mpc, but may be quite significant if using the
information at  larger $k$.  Since the FoG effect is still very difficult
to reliably model from first principles, the empirical approach
we developed in this paper will be useful for correcting the residual
FoG effect, or at least making a sanity check of the residual FoG
effect. 

Recently \citet{Masakietal:12} developed a method to generate a
mock catalog of the SDSS LRGs based on the abundance matching method,
where the central and satellite subhalos in $N$-body simulation output
are identified as the places hosting LRGs until the number density of
the matched subhalos is similar to that of SDSS LRGs. They nicely showed
that the mock catalog fairly well reproduces the SDSS measurements for
the satellite LRG fraction, the projected correlation function, the LRG
weak lensing, and the redshift-space power spectrum. The mock catalog
predicts that some of LRGs in the single-LRG systems are off-centered.
Also interestingly, they found that off-centered LRGs in the
multiple-LRG halos typically have a velocity dispersion of $500~{\rm
km/s}$, which is consistent with our finding. It would be worth further
exploring a study to compare 
 the measurement with 
the mock
catalog and the halo model in order to understand the nature of LRGs.

There are some limitations in the FoG correction for the SDSS LRG
catalog. First, the statistical accuracy of the LRG-galaxy lensing
measurement is limited by the small number density of background
galaxies (about 1.2 arcmin$^{-2}$ for the SDSS photometric galaxies),
and also by the typical SDSS seeing  (typically
1.2$^{\prime\prime}$ FWHM).  Second, we lack a 
good indicator of the halo center for each single LRG
system.
If deeper imaging data is available, it may help us to
identify member galaxies around each LRG 
and then estimate the richness of each LRG
system, e.g. by using optical richness or the number of member
galaxies. In addition, since LRGs are selected by a well-tuned
color cut, there may be more luminous galaxy(ies) in
some of LRG systems, e.g., blue-color brightest cluster galaxy. If such
richness information on member galaxies or other luminous galaxies are
available, we can use the method in this paper for all the
LRG systems by comparing the correlation measurements for the different
halo center proxies. 
Then, we may be able to correct the residual
FoG effect more accurately for all the LRG systems. 
The Subaru HSC
survey is planning to have an overlap with the SDSS region, for about
1500 square degrees. The number density of galaxies behind $z\sim 0.3$
LRGs, usable for the shape measurement analysis, is about 20
arcmin$^{-2}$, a factor of almost 20 larger than for 
SDSS. Although the area of the HSC survey is
about a factor of 7 smaller than that of the SDSS (about 10000 deg$^2$),
the HSC can improve the measurement accuracies by more
than a factor of $\sqrt{20/7}=1.7$ at each radial bin, due to the additional
gain in the lensing efficiency based on the higher mean redshift of source
galaxies than in the SDSS. 
Since LRGs in the 1500 sq. degrees area can be
safely considered as a fair, representative sample of LRGs, the HSC
survey would improve the accuracy of the FoG correction for the entire
SDSS LRG sample, and will more
generally increase our understanding of the nature of LRGs and their
relation to their 
host halos. Furthermore, the higher number density of imaging
galaxies can also improve the accuracies of the projected
cross-correlation measurements.

The deeper depth of an imaging survey such as the HSC survey is also
very useful to extend the method of this paper to massive galaxies at higher
redshifts, which is being surveyed by the SDSS-III BOSS survey.  BOSS has
a denser sampling of massive galaxies by a factor of 3-4 compared to the SDSS, so
the FoG contamination can be more significant. Again, by having
background galaxies around each BOSS galaxy and/or photometric galaxies in
the BOSS galaxy redshift range ($z=[0.4,0.75]$), we can use  weak lensing
and cross-correlation measurements to observationally explore the nature
of those massive red galaxies and to remove the FoG contamination for doing
cosmology. These are all crucial in order to reliably use the amplitude
and shape information of the galaxy power spectrum in order
to constrain the growth rate as well as
cosmological parameters such as the neutrino masses. 

Our paper demonstrates the complementarity of spectroscopic
and imaging surveys for observationally constraining the connection
between galaxies and dark matter halos. This is just one example, and
there are various synergies available, if the imaging and spectroscopic
surveys see the {\em same} region of the sky (or have at least a
sufficiently large overlapping area). This is the case for the SDSS BOSS
and the HSC survey, the HSC and PFS surveys, and the Euclid survey.
First, as we stressed in this paper, we can measure the galaxy-galaxy
lensing or the cross-correlation as a function of the projected radius
$R$ in the comoving unit from the spectroscopic galaxies, rather than
the angular separation. The correlations do not mix different scales at
different redshifts, and can preserve the same physical radius even if
taking a wider redshift slice in the correlation analysis, e.g., the
virial radii of halos or the BAO scales. Hence, such correlation
functions measured against the projected radii $R$
yield a measurements of the three-dimensional power spectrum
or correlation functions (see Eqs.~\ref{eq:lens_2h_app} and
\ref{eq:wR_app}).  This is obvious, but has not been explicitly stressed
in the literature. Furthermore, combining the imaging and
spectroscopic surveys allows for a reconstruction of galaxy bias
functions on large scales as well as a calibration of other systematic
errors such as the halo mass, photo-$z$ errors, and shear multiplicative
biases  
\citep{Reyesetal:10,Baldaufetal:10,OguriTakada:11,Mandelbaumetal:12}. Thus
various synergies between imaging and spectroscopic surveys need to be
more carefully explored in order to attain the full potential of these
surveys for deriving the most stringent and robust constraints possible.

\bigskip
 
\section*{Acknowledgments}
We thank Shirley Ho, Issha Kayo, Surhud More, Beth Reid, Uros Seljak
and Kazuhiro Yamamoto for useful discussion and valuable comments.
We also acknowledge an anonymous referee for useful and
  constructive comments that helped improve the manuscript of our
  paper.  CH acknowledges support from a Japan Society for Promotion
of Science (JSPS) fellowship green and Grant-in-Aid for
  Scientific Research from the Ministry of Education, Science, Sports,
  and Culture, Japan, No. 24740160.  RM acknowledges the support of
the Department of Energy Early Career Award program for some of the
duration of this work. MT greatly thanks Department of Astrophysical
Sciences, Princeton University for its warm hospitality during his
visit, where this work was done. MT and RM also thank
 the Aspen Center for Physics and the NSF Grant \#1066293 for
their warm hospitality, where this work was partly done. 
 DNS and CH acknowledge support from
NSF grant AST-0707731 and the NASA AST theory program.  DNS thanks the
IPMU for its warm hospitality during his visit, where the work was
completed.  This work is in part supported in part by JSPS
Core-to-Core Program ``International Research Network for Dark
Energy'', by Grant-in-Aid for Scientific Research from the JSPS
Promotion of Science, by Grant-in-Aid for Scientific Research on
Priority Areas No. 467 ``Probing the Dark Energy through an Extremely
Wide \& Deep Survey with Subaru Telescope'', by World Premier
International Research Center Initiative (WPI Initiative), MEXT,
Japan, by the FIRST program ``Subaru Measurements of Images and
Redshifts (SuMIRe)'', CSTP, Japan, and by the exchange program between
JSPS and DFG.

\appendix

\section{Impact of an NFW profile assumption on the off-centering
parameters from the LRG-galaxy cross-correlation}
\label{sec:genNFW}

\begin{figure}
\begin{center}
\includegraphics[width=14cm]{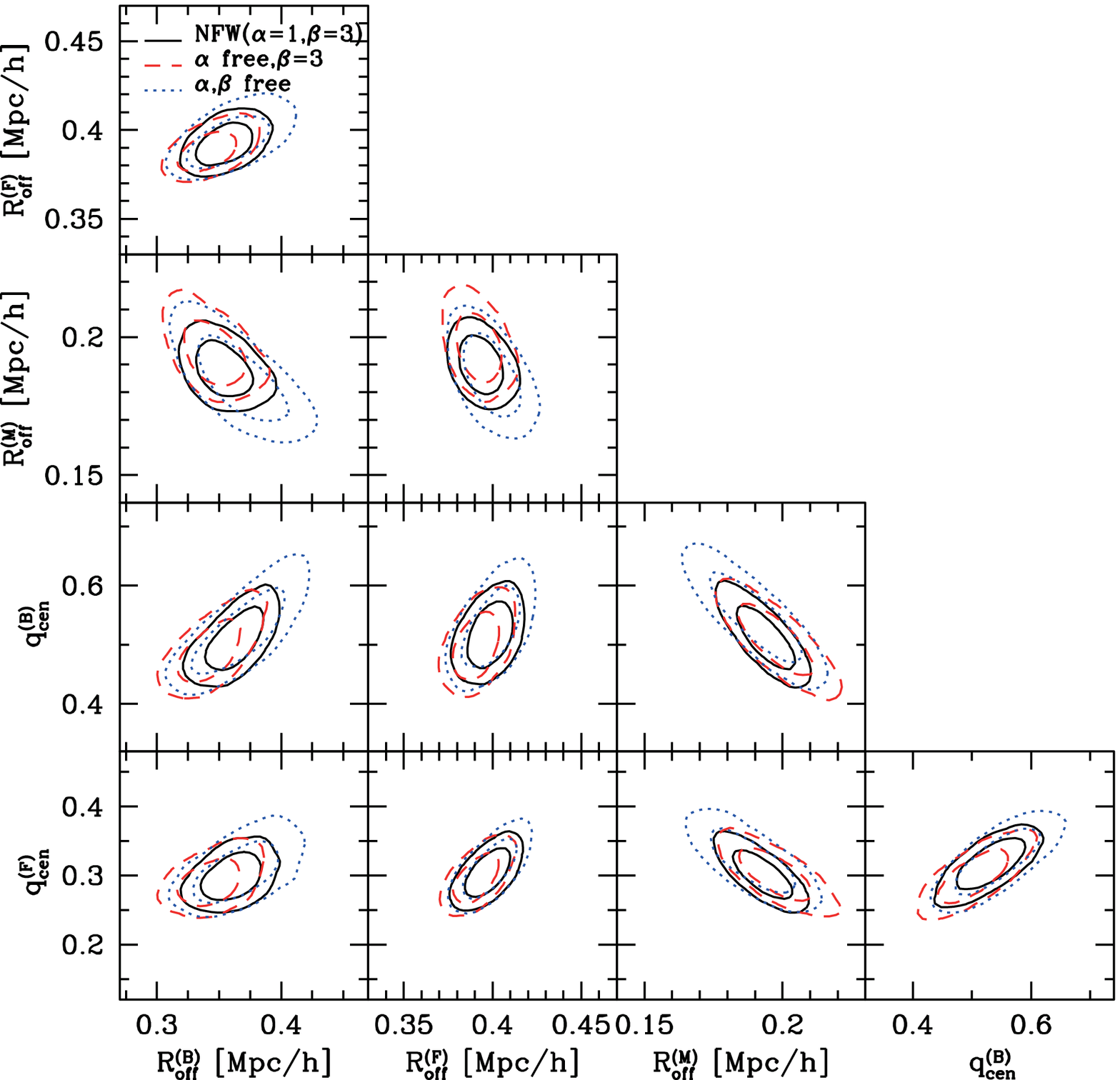}
\caption{Similarly to Fig.\ref{fig:2dpost_angcor}, but the
marginalized constraint contours (68 and 95 per cent C.L.) derived from
the measured cross-correlations of LRGs with photometric galaxies for
the different halo centers (the right panel of Fig.\ref{fig:wl-wR_obs}),
where we used the generalized NFW profile, $n_g(r)\propto
1/[r^\alpha(1+r/r_s)^{\beta-\alpha}]$, for the radial distribution of
photometric galaxies and allow the two parameters ($\alpha,\beta$) to
freely vary in the model fitting in addition to the set of parameters
(Eq.~\ref{eq:paras_wR}). Note that we used the range of $0.1\le R\le
4~{\rm Mpc}/h$, and the solid contours are the same as those in
Fig.\ref{fig:2dpost_angcor}. The dashed contours are the results when
varying only $\alpha$, but fixing $\beta=3$. The dotted contours are the
results when varying both $\alpha$ and $\beta$, where we employed the
priors $\alpha=[0,3]$ and $\beta=[1,5]$. The best-fit parameters
 $\alpha=1.4\pm 0.2$ and $\beta=3.3\pm 0.2$.
The confidence contours are
only slightly enlarged for the generalized NFW profile, and the results
are consistent with the results obtained by assuming an NFW profile. 
Thus the results show that the off-center parameters are well
 constrained, irrespective of the form of the radial profile, by
 combining the measurements of different centers. 
 \label{fig:genNFW} }
\end{center}
\end{figure}

One of the main uncertainties in the off-center constraints we derived
 from the measured cross-correlation functions of LRGs with
 photometric galaxies is the
 assumption of an NFW profile for the photometric galaxies. Here we
 study how the off-centering parameters are changed by relaxing the NFW
 assumption. 
To study this, we assume that the radial distribution of photometric
galaxies follows a generalized NFW profile given as 
$n_g(r)\propto 1/[r^\alpha (1+r/r_s)^{\beta-\alpha}]$, 
where $r_s$ is the scale radius
given in terms of the virial radius and the halo concentration as 
$r_s=r_{180b}/c_{180b}$. In the model fitting, we allow the two
parameters ($\alpha,\beta$) to freely vary, in addition to the set of
free parameters in Eq.~(\ref{eq:paras_wR}). We employ flat 
priors on the new parameters as $\alpha=[0,3]$ and $\beta=[1,5]$, where
the profile with $\alpha=1 $ and $\beta=3$ corresponds to an NFW
profile. 

Fig.\ref{fig:genNFW} shows the marginalized confidence regions for the
off-centering parameters, similarly to Fig.\ref{fig:2dpost_angcor}, where
we used the cross-correlations for the different centers in the range of
$0.1<R<4~{\rm Mpc}/h$. The dashed contours are the results when we allow
only $\alpha$ to freely vary, and the dotted contours are the results
when both $\alpha$ and $\beta$ are freely varied.  The best-fit
parameters are $\alpha=1.4\pm 0.2$ and $\beta=3.3\pm 0.2$ for the latter
model, and the best-fit model is consistent with an NFW profile.  These
results are compared to the solid contours, which are the results in
Fig.\ref{fig:2dpost_angcor}. The figure shows that, even if allowing a
more generalized profile, 
each of the off-centering parameters 
is well constrained
 and the confidence region is only slightly
enlarged by adding the two 
parameters.
Thus the degeneracy between the
off-centering parameters and the profile parameters is efficiently broken
by comparing the cross-correlations for the different choices of halo centers.
As we discussed in the main text (see Section~\ref{sec:fog_residual}), 
the off-centering parameters for the single-LRG systems 
are
significantly degraded if relaxing the NFW assumption, because it does
not allow for comparison between the measurements of different centers
(only the single LRG itself can
be taken as the center in the correlation calculation): we found 
$q_{\rm cen}=0.76\pm 0.06$ and 
$R_{\rm off}=(0.26\pm 0.03)~{\rm Mpc}/h$ for the NFW assumption, while
these constraints became 
 $q_{\rm cen}=0.79\pm 0.08$ and $R_{\rm
off}=0.26\pm 0.15~{\rm Mpc}/h$ for the generalized NFW profile.  Thus
the constraints on the offset radius are considerably weaker.

\section{Dependence of the off-centering effects on the magnitude
 difference between BLRG and FLRG}
\label{sec:lrg_dm}

\begin{figure}
\begin{center}
\includegraphics[width=8.5cm]{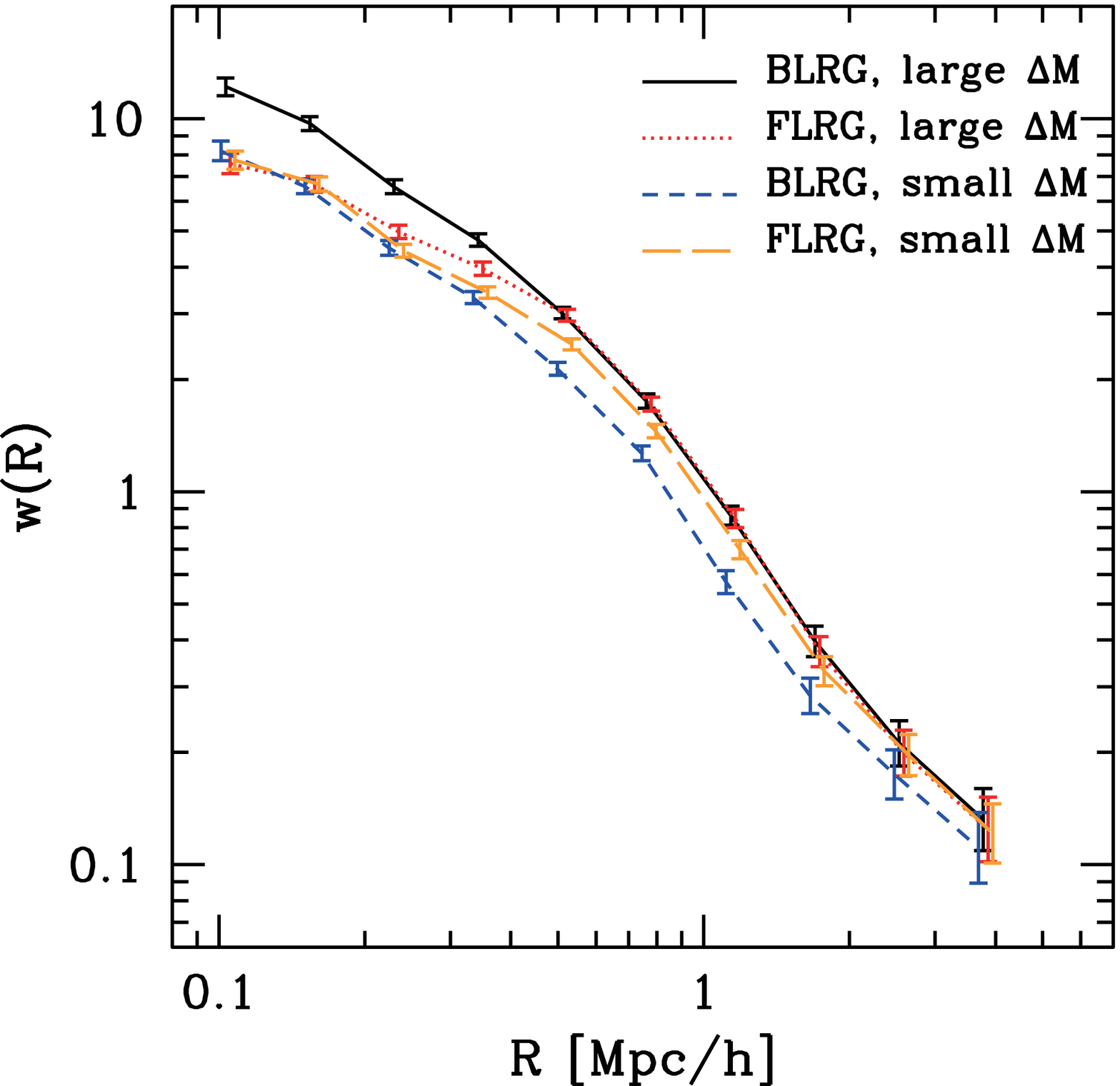}
\includegraphics[width=8.5cm]{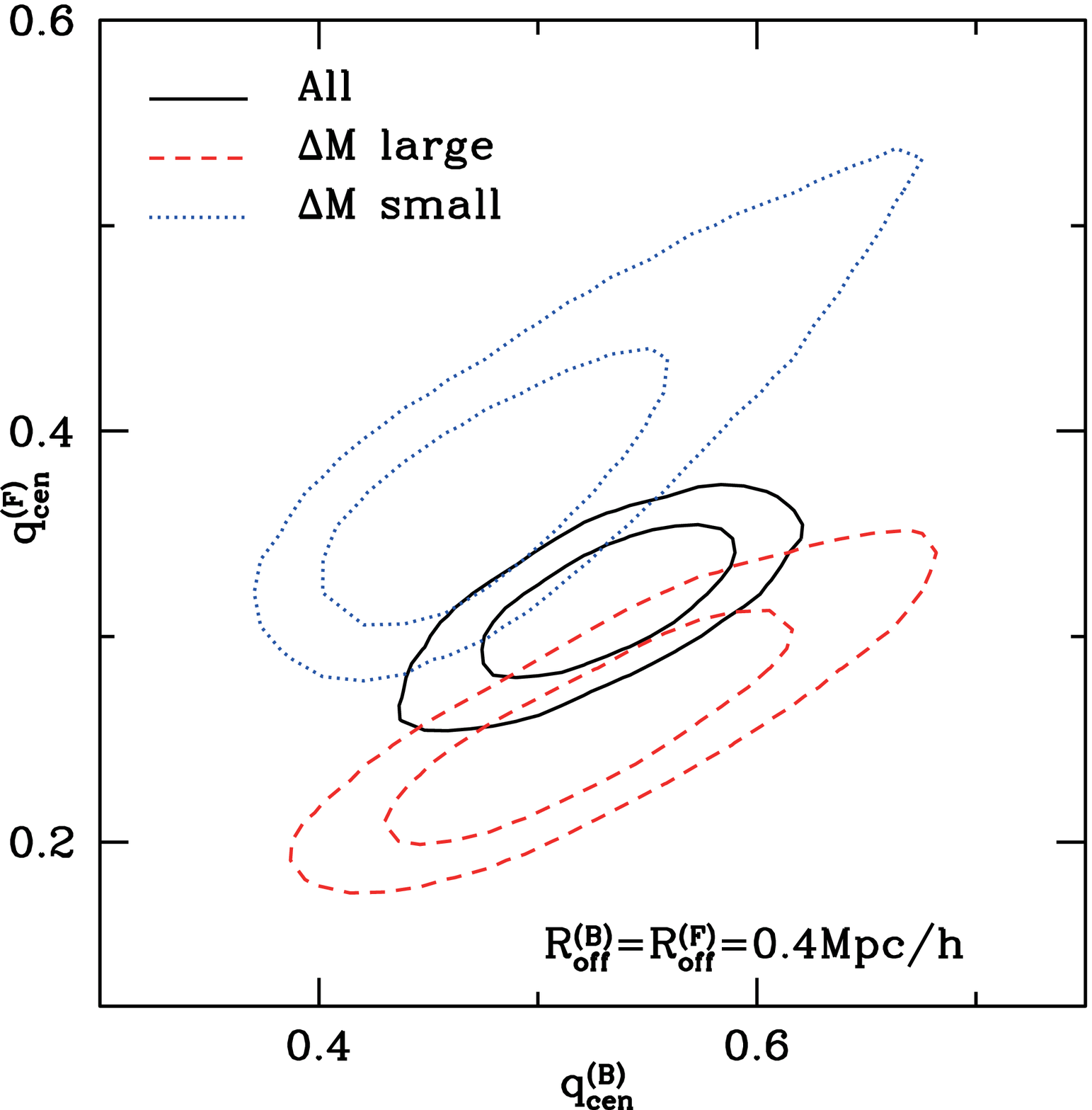}
\caption{{\em Left panel}: the cross-correlation functions of the BLRG
or FLRG position with the photometric galaxies, for the two subsamples
of the multiple-LRG systems, where the multiple-LRG systems are divided
into halves. One subsample is defined by the systems which
have the larger magnitude difference between BLRG and FRLG than the
median (larger $\Delta M$), while the other contains the remaining systems, 
those 
 with the smaller magnitude difference (smaller $\Delta M$). For each
 subsample, we measured the cross-correlations for the BLRG or FLRG
 centers as indicated by the legend. Note that the correlation
 amplitudes are not reliable due to the dilution effect caused by
 photo-$z$ errors. {\em Right panel}: the best-fit off-centering parameters
 obtained from each sub-sample. Since we cannot obtain sufficiently
 tight constraints on the parameters
due to insufficient statistical
 power, we fixed the typical off-centering radius to $R=400~{\rm
 kpc}/h$ for both the BLRG and FLRG centers, and constrained the 
 fraction of central BLRGs or FLRGs, $q_{\rm cen}^{\rm BLRG}$ or $q_{\rm
 cen}^{\rm FLRG}$. The fraction of the central BLRGs
is not different 
 for the two subsamples, but the fraction of the central FLRGs is
 smaller for the subsample with the larger magnitude difference.  
\label{fig:wr_dm} }
\end{center}
\end{figure}

We found that some BLRGs in the multiple-LRG systems are satellite
galaxies. As shown in Fig.\ref{fig:LRGs_dm}, the BLRG in some systems is
brighter than FLRG by only a few tenths of a magnitude, implying that BLRG
and FLRG are not so different. Given this fact, one might consider that
satellite BLRGs are mainly from such multiple-LRG systems where
the FLRG is
instead a central galaxy and the BLRG is a satellite. On the other hand, for
systems where the BLRG and FLRG luminosities are significantly
  different, the BLRG may have
a higher chance to be central. In this appendix, we study this
possibility by
dividing the multiple-LRG systems into halved subsamples in terms of the
magnitude difference between BLRG and FLRG in each system. One halved
subsample contains the systems for which the BLRG and FLRG have a smaller
magnitude difference than the median (the ``small $\Delta$M''
subsample), while the other subsample contains the remaining systems
(the larger ``$\Delta$M''). 

The left panel of Fig.\ref{fig:wr_dm} shows the measured
cross-correlations for the halved subsamples, using either BLRG or FLRG
centers. The cross-correlations for the different subsamples and centers
differ from each other, especially in their amplitudes. However, as we
stressed in Section~\ref{sec:angcor}, we cannot directly compare the
correlation amplitudes because the amplitude suffers from a dilution due
to photo-$z$ errors. Instead we need to use the shape of the correlation
function to infer the off-centering effects. By fitting the halo model to
the cross-correlations, we estimate the off-centering parameters. However,
due to the insufficient statistical accuracies of the halved subsample,
we were not able to obtain tight constraints on the parameters, after
marginalizing over other parameters. Hence, in the right panel, we show
the constraints on the fractions of central BLRGs or FLRGs for each of
the two subsamples when fixing the off-centering radius to $0.4~{\rm
Mpc}/h$, the best-fit value for all the multiple-LRG systems. The plot
shows that the fraction of central BLRGs is similar for the two
subsamples, $q_{\rm cen}^{\rm BLRG}\simeq 0.5$, but the fraction of
central FLRGs is higher for the subsample with the smaller magnitude
difference between BLRG and FLRG than the other subsample, supporting
the hypothesis that the FLRG is not so different from
the BLRG, and the FLRG is instead a central
galaxy in the subsample. 

\section{Implication for the residual FoG effect in the LRG power
spectrum from the LRG-galaxy weak lensing}
\label{sec:fog_wl}

\begin{figure}
\begin{center}
\includegraphics[width=10.6cm]{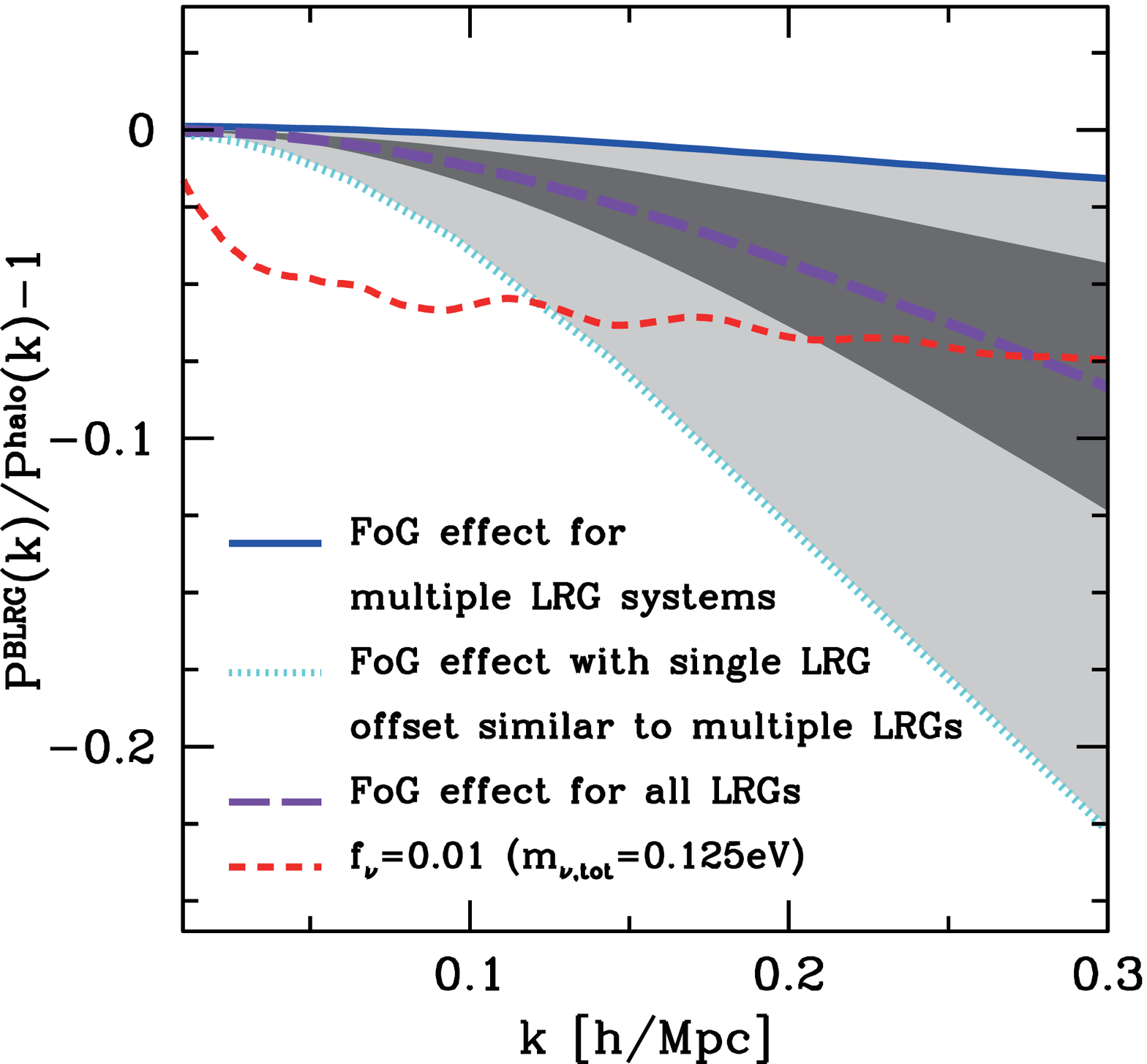}
\caption{
Similarly to Fig.\ref{fig:pk_fog}, but the residual FoG effect
 implied from the off-center parameters constrained from the LRG-galaxy
 weak lensing for the single-LRG systems (the left panel of 
Fig.\ref{fig:obs_single}). 
\label{fig:fog_wl}
}
\end{center}
\end{figure}

 In Section~\ref{sec:fog_residual}, we discussed a possible
residual FoG effect in the LRG redshift-space power spectrum arising
from the off-centering effects of LRGs. To discuss this effect, we used
the cross-correlation measurement for the single-LRG systems, and 
the constraints on off-centering parameters to infer a possible FoG
effect in Fig.\ref{fig:pk_fog}. For completeness,
Fig.\ref{fig:fog_wl} shows the residual FoG effect 
 implied from the off-centering of LRG-galaxy weak
  lensing (assuming the NFW profile).  Due to the smaller
signal-to-noise ratio for the LRG-galaxy compared to the galaxy 
cross-correlation, the off-centering parameters
are not well constrained; $q_{\rm cen}= 0.76\pm 0.18 $ and $R_{\rm
  off}=0.38\pm 0.25$Mpc/h  for single-LRG systems and
  $q_{\rm cen}= 0.63\pm 0.21 $ and $R_{\rm off}=0.44\pm 0.22$Mpc/h for
  multiple-LRG systems. The dark shaded region in
Fig.\ref{fig:fog_wl} shows the implied range of residual FoG effect,
which is wider than that in Fig.~\ref{fig:pk_fog}.  Nevertheless, we
should emphasize that the 
FoG effects in
Figs.~\ref{fig:pk_fog} and \ref{fig:fog_wl} are consistent with each
other within the error bars.

\bibliographystyle{mn2e} \bibliography{mn-jour,refs} \label{lastpage}
\end{document}